\font\fiverm=cmr5
\font\fivebf=cmbx5
\font\fivei=cmmi5

\font\fivesy=cmsy5

\font\sevenrm=cmr7
\font\sevenbf=cmbx7
\font\seveni=cmmi7

\font\sevensy=cmsy7

\font\ninerm=cmr9
\font\ninebf=cmbx9

\font\nineit=cmmi9
\font\ninei=cmmi9 
\font\ninesy=cmsy9  
\font\nineex=cmex10
\font\tenrm=cmr10
\font\tenbf=cmbx10
\font\tensl=cmsl10
\font\tenit=cmmi10
\font\teni=cmmi10 
 
\font\tensy=cmsy10
\font\tenex=cmex10
\font\twelverm=cmr12
\font\twelvebf=cmbx12
\font\twelvesl=cmsl12
\font\twelveit=cmmi12
\font\twelvei=cmmi12
\font\twelvesy=cmsy10 scaled\magstep1


%
%
 %
%
 \def\ninepoint{%
   \normalbaselineskip=11pt
   \def\rm{\fam0\ninerm}%
   \def\it{\fam0\nineit}%
   \def\bf{\fam\bffam\ninebf}%
   \def\bi{\fam\bffam\ninebf}%
   \def\rmit{\fam0\ninerm\def\it{\fam0\nineit}}%
   \def\bfit{\fam\bffam\ninebf\def\it{\bi}}%
   \def\bsl{\fam\bffam\ninebsl}
   \textfont0=\ninerm\scriptfont0=\sevenrm\scriptscriptfont0=\fiverm
   \textfont1=\ninei\scriptfont1=\seveni\scriptscriptfont1=\fivei
   \textfont2=\ninesy\scriptfont2=\sevensy\scriptscriptfont2=\fivesy
   \textfont3=\nineex \scriptfont3=\tenex \scriptscriptfont3=\tenex
   \textfont\bffam=\ninebf\scriptfont\bffam=\sevenbf\scriptscriptfont\bffam=
     \fivebf
   \normalbaselines\rm}%
%
 \def\tenpoint{%
   \normalbaselineskip=12pt
   \def\rm{\fam0\tenrm}%
   \def\it{\fam0\tenit}%
   \def\bf{\fam\bffam\tenbf}%
   \def\bi{\fam\bffam\tenbf}%
   \def\rmit{\fam0\tenrm\def\it{\fam0\tenit}}%
   \def\bfit{\fam\bffam\tenbf\def\it{\bi}}%
   \def\bsl{\fam\bffam\tenbsl}
   \textfont0=\tenrm\scriptfont0=\sevenrm\scriptscriptfont0=\fiverm
   \textfont1=\teni\scriptfont1=\seveni\scriptscriptfont1=\fivei
   \textfont2=\tensy\scriptfont2=\sevensy\scriptscriptfont2=\fivesy
   \textfont3=\tenex \scriptfont3=\tenex \scriptscriptfont3=\tenex
     \fivebf
   \textfont\bffam=\tenib\scriptfont\bffam=\sevenib\scriptscriptfont\bffam=
     \fiveib
   \normalbaselines\rm}%
%

%

     
\font\twelverm=cmr10 scaled 1200    \font\twelvei=cmmi10 scaled 1200
\font\twelvesy=cmsy10 scaled 1200   \font\twelveex=cmex10 scaled 1200
\font\twelvebf=cmbx10 scaled 1200   \font\twelvesl=cmsl10 scaled 1200
\font\twelvett=cmtt10 scaled 1200   \font\twelveit=cmti10 scaled 1200
     
\skewchar\twelvei='177   \skewchar\twelvesy='60
     
     
\def\twelvepoint{\normalbaselineskip=12.4pt
  \abovedisplayskip 12.4pt plus 3pt minus 9pt
  \belowdisplayskip 12.4pt plus 3pt minus 9pt
  \abovedisplayshortskip 0pt plus 3pt
  \belowdisplayshortskip 7.2pt plus 3pt minus 4pt
  \smallskipamount=3.6pt plus1.2pt minus1.2pt
  \medskipamount=7.2pt plus2.4pt minus2.4pt
  \bigskipamount=14.4pt plus4.8pt minus4.8pt
  \def\rm{\fam0\twelverm}          \def\it{\fam\itfam\twelveit}%
  \def\sl{\fam\slfam\twelvesl}     \def\bf{\fam\bffam\twelvebf}%
  \def\mit{\fam 1}                 \def\cal{\fam 2}%
  \def\tt{\twelvett}
  \textfont0=\twelverm   \scriptfont0=\tenrm   \scriptscriptfont0=\sevenrm
  \textfont1=\twelvei    \scriptfont1=\teni    \scriptscriptfont1=\seveni
  \textfont2=\twelvesy   \scriptfont2=\tensy   \scriptscriptfont2=\sevensy
  \textfont3=\twelveex   \scriptfont3=\twelveex  \scriptscriptfont3=\twelveex
  \textfont\itfam=\twelveit
  \textfont\slfam=\twelvesl
  \textfont\bffam=\twelvebf \scriptfont\bffam=\tenbf
  \scriptscriptfont\bffam=\sevenbf
  \normalbaselines\rm}
     
     
\def\tenpoint{\normalbaselineskip=12pt
  \abovedisplayskip 12pt plus 3pt minus 9pt
  \belowdisplayskip 12pt plus 3pt minus 9pt
  \abovedisplayshortskip 0pt plus 3pt
  \belowdisplayshortskip 7pt plus 3pt minus 4pt
  \smallskipamount=3pt plus1pt minus1pt
  \medskipamount=6pt plus2pt minus2pt
  \bigskipamount=12pt plus4pt minus4pt
  \def\rm{\fam0\tenrm}          \def\it{\fam\itfam\tenit}%
  \def\sl{\fam\slfam\tensl}     \def\bf{\fam\bffam\tenbf}%
  \def\smc{\tensmc}             \def\mit{\fam 1}%
  \def\cal{\fam 2}%
  \textfont0=\tenrm   \scriptfont0=\sevenrm   \scriptscriptfont0=\fiverm
  \textfont1=\teni    \scriptfont1=\seveni    \scriptscriptfont1=\fivei
  \textfont2=\tensy   \scriptfont2=\sevensy   \scriptscriptfont2=\fivesy
  \textfont3=\tenex   \scriptfont3=\tenex     \scriptscriptfont3=\tenex
  \textfont\itfam=\tenit
  \textfont\slfam=\tensl
  \textfont\bffam=\tenbf \scriptfont\bffam=\sevenbf
  \scriptscriptfont\bffam=\fivebf
  \normalbaselines\rm}
     

{\obeylines\gdef\
{}}
\def\singlespace{\baselineskip=\normalbaselineskip}
\def\oneandathirdspace{\baselineskip=\normalbaselineskip
  \multiply\baselineskip by 4 \divide\baselineskip by 3}

\def\doublespace{\baselineskip=\normalbaselineskip \multiply\baselineskip by 2}

\newcount\firstpageno
\firstpageno=2
\footline={\ifnum\pageno<\firstpageno{\hfil}\else{\hfil\twelverm\folio\hfil}\fi}
\let\rawfootnote=\footnote              
\def\footnote#1#2{{\rm\singlespace\parindent=0pt\rawfootnote{#1}{#2}}}

     
\hsize=6.5truein
\hoffset=0truein
\vsize=8.9truein
\voffset=0truein
\parskip=\medskipamount
\twelvepoint            
\doublespace            
\overfullrule=0pt       
     
     
\def\preprintno#1{
 \rightline{\rm #1}}    
     
     
\def\ref#1{Ref. #1}                     

\def\frac#1#2{{\textstyle{#1 \over #2}}}
\def\half{{\textstyle{ 1\over 2}}}

\def\sla{\raise.15ex\hbox{$/$}\kern-.57em}
\def\leaderfill{\leaders\hbox to 1em{\hss.\hss}\hfill}
\def\twiddle{\lower.9ex\rlap{$\kern-.1em\scriptstyle\sim$}}
\def\bigtwiddle{\lower1.ex\rlap{$\sim$}}
\def\gtwid{\mathrel{\raise.3ex\hbox{$>$\kern-.75em\lower1ex\hbox{$\sim$}}}}
\def\ltwid{\mathrel{\raise.3ex\hbox{$<$\kern-.75em\lower1ex\hbox{$\sim$}}}}
\def\square{\kern1pt\vbox{\hrule height 1.2pt\hbox{\vrule width 1.2pt\hskip 3pt
   \vbox{\vskip 6pt}\hskip 3pt\vrule width 0.6pt}\hrule height 0.6pt}\kern1pt}

\def\tablerule{\tablespace\noalign{\hrule}\tablespace}

\def\comp{{\rm C}\llap{\vrule height7.1pt width1pt depth-.4pt\phantom t}}

\def\Fint{\rlap{$\Biggl\rfloor$}\Biggl\lceil}

\def\m@th{\mathsurround=0pt }
\def\leftrightarrowfill{$\m@th \mathord\leftarrow \mkern-6mu
 \cleaders\hbox{$\mkern-2mu \mathord- \mkern-2mu$}\hfill
 \mkern-6mu \mathord\rightarrow$}
\def\overleftrightarrow#1{\vbox{\ialign{##\crcr
     \leftrightarrowfill\crcr\noalign{\kern-1pt\nointerlineskip}
     $\hfil\displaystyle{#1}\hfil$\crcr}}}

\input psfig.sty
\singlespace
\preprintno{hep-ph/9602317}
\preprintno{CPTH-S421.1295}
\preprintno{CRETE-96-11}
\preprintno{UFIFT-HEP-96-4}
\preprintno{Revised June, 1996}
\vskip 2cm
\centerline{\bf One Loop Graviton Self-Energy In A Locally De Sitter Background}
\vskip 2cm
\centerline{\bf N. C. Tsamis$^{*}$}
\vskip .5cm
\centerline{\it Centre de Physique Th\'eorique, Ecole Polytechnique}
\centerline{\it Palaiseau 91128, FRANCE}
\vskip .5cm
\centerline{and}
\vskip .5cm
\centerline{\it Theory Group, FO.R.T.H.}
\centerline{\it Heraklion, Crete 71110, GREECE}
\vskip 1cm
\centerline{and}
\vskip 1cm
\centerline{\bf R. P. Woodard$^{\dagger}$}
\vskip .5cm
\centerline{\it Department of Physics, University of Florida}
\centerline{\it Gainesville, FL 32611, USA}
\vskip 2cm
\centerline{ABSTRACT}
\itemitem{}{\tenpoint The graviton tadpole has recently been computed at two
loops in a locally de Sitter background. We apply intermediate results of this 
work to exhibit the graviton self-energy at one loop. This quantity is 
interesting both to check the accuracy of the first calculation and to 
understand the relaxation effect it reveals. In the former context we show that
the self-energy obeys the appropriate Ward identity. We also show that its flat
space limit agrees with the flat space result obtained by Capper in what should
be the same gauge.}
\footnote{}{$^*$~~e-mail: tsamis@iesl.forth.gr and 
tsamis@orphee.ploytechnique.fr}
\footnote{}{$^{\dagger}$~~e-mail: woodard@phys.ufl.edu}

\vfill\eject
 
\doublespace

\centerline{I. INTRODUCTION}

We have suggested that inflation ended in the early universe because the 
quantum gravitational back-reaction slowly generated negative vacuum energy 
which eventually screened a not unnaturally small, positive cosmological 
constant.$^{1,2}$ This is an attractive scenario for solving the problem of the 
cosmological constant because:

{\oneandathirdspace
\item{(1)} It operates in the far infrared where general relativity can be used
reliably as a quantum theory of gravitation;
\item{(2)} It introduces no new light quanta which would embarrass low energy
phenomenology;
\item{(3)} It has the potential to make unique predictions because gravity is
the only phenomenologically viable theory which possesses the essential feature
of massless quanta whose self-interactions are not conformally invariant; and
\item{(4)} The weakness of gravitational interactions makes the process slow 
enough to account for a long period of inflation.

}

\doublespace
\noindent If this proposal is correct there will be far-reaching consequences 
for theories of the very early universe. Scarcely less significant, in the long
run, is the fact that contact will finally have been made between observed 
reality and the hitherto murky realm of quantum gravity.

We have recently done a calculation which establishes the validity of our
scenario for at least as long as perturbation theory remains reliable. The 
quantity we computed is the expectation value of the invariant element, 
starting from a homogeneous and isotropic, locally de Sitter, free vacuum on 
the manifold $T^3 \times \Re$:
$$\Bigl\langle \Omega \Bigl\vert \; g_{\mu \nu}(t,{\vec x}) \;
dx^{\mu} dx^{\nu} \; \Bigr\vert \Omega \Bigr\rangle = - dt^2 + 
{\rm a}^2(t) \; d{\vec x} \cdot d{\vec x} \eqno(1.1)$$
The rate of spacetime expansion is measured using the coordinate invariant 
effective Hubble constant:
$$H_{\rm eff}(t) \equiv {1 \over {\rm a}(t)} {d {\rm a}(t) \over dt} 
\eqno(1.2)$$
One loop tadpoles make no contribution because they are ultra-local whereas
infrared effects derive from the causal and coherent superposition of
interactions throughout the past lightcone. The first secular effect comes from
the two loop diagrams shown in Fig.~1. At the end of a very long calculation we
obtain the following result:$^{3,4}$
$$H_{\rm eff}(t) = H \Biggl\{1 - \Bigl({\kappa H \over 4 \pi}\Bigr)^4 \Bigl[
\frac16 (Ht)^2 + {\cal O}(Ht)\Bigr] - {\cal O}(\kappa^6)\Biggr\} 
\eqno(1.3)$$
where $H \equiv \sqrt{\frac13 \Lambda}$ is the Hubble constant at the onset of
inflation and $\kappa^2 \equiv 16 \pi G$ is the usual loop counting parameter
of perturbative quantum gravity. We have also been able to show that the $\ell$
loop contribution to the bracketed term can be no stronger than $- \# (\kappa 
H)^{2\ell} (Ht)^{\ell}$.

\vskip -1.5cm

\centerline{\psfig{figure=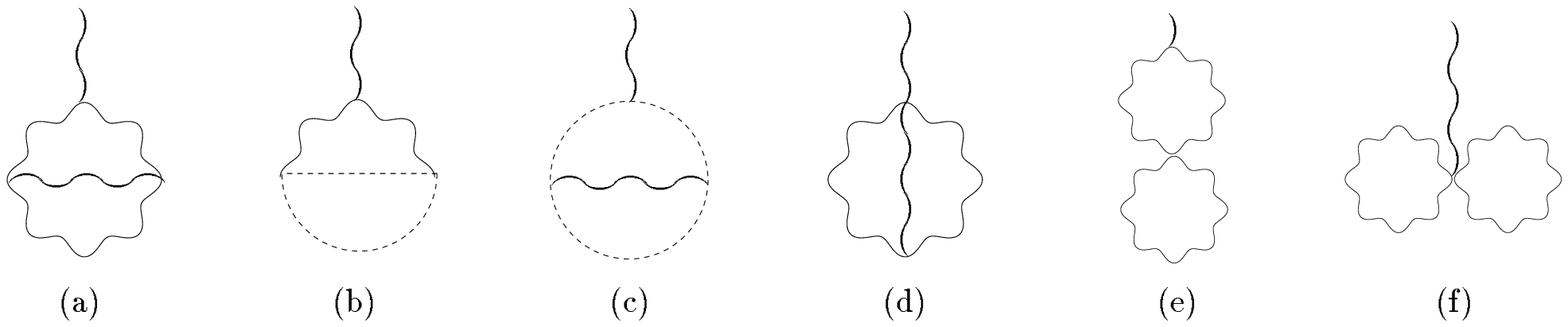,height=21cm,bbllx=0bp
,bblly=0bp,bburx=596bp,bbury=843bp,rheight=5.5cm,rwidth=12.7cm}}

{\bf Fig.~1:} {\ninepoint Two-loop contributions to the background 
geometry. Gravitons reside on wavy lines and ghosts 
\vskip -8pt \noindent \hglue 2.40truecm on segmented lines.}

\vskip 0.5cm

It is not easy to compute at two loops even for scalar field theories on flat
space, and truth can sometimes remain well hidden amidst the forest of indices 
which characterize any calculation in quantum gravity. In fact only one other 
two loop result has been obtained for quantum gravity, and this was limited to 
the ultraviolet divergent part of the standard, in-out effective action for 
zero cosmological constant.$^5$ To study the ultraviolet one can use asymptotic
expansions in which the effects of spacetime curvature are segregated from what
is basically a calculation in flat space. The infrared does not allow this 
simplification; we had to obtain the full propagators on a curved background
and integrate them against the appropriate interaction vertices over a large
invariant spacetime volume. There was an additional complication in having to 
use Schwinger's formalism$^6$ to obtain a true expectation value rather than an 
in-out matrix element. One naturally wonders, therefore, about the accuracy of 
a result such as the one we are reporting. This concern is heightened by the 
fact that so much of the relevant formalism has only recently been developed.

The possibility for dramatic checks on the consistency of the formalism and on
our proficiency in applying it is provided by the manner in which we computed 
the two most complicated diagrams, (1a) and (1b). In order to economize on the 
size of intermediate expressions we evaluated the lower loops first and then 
contracted them into the two upper propagators and the final vertex as 
represented diagramatically in Fig.~2. A consequence is that we can extract the
one loop graviton self-energy. This quantity can be subjected to two powerful 
tests: the flat space limit ($H \rightarrow 0$ with $\kappa$ and $t$ held fixed)
and the Ward identity. The Ward identity checks our gauge fixing procedure, our
solution for the ghost and graviton propagators, our 3-point vertices, and the
automated reduction procedures through which we contracted propagators into
vertices and acted the various derivatives. In addition to providing a largely
complementary check on all these things, the flat space limit tests the overall
proportionality constant.

\vskip 1cm

\centerline{\psfig{figure=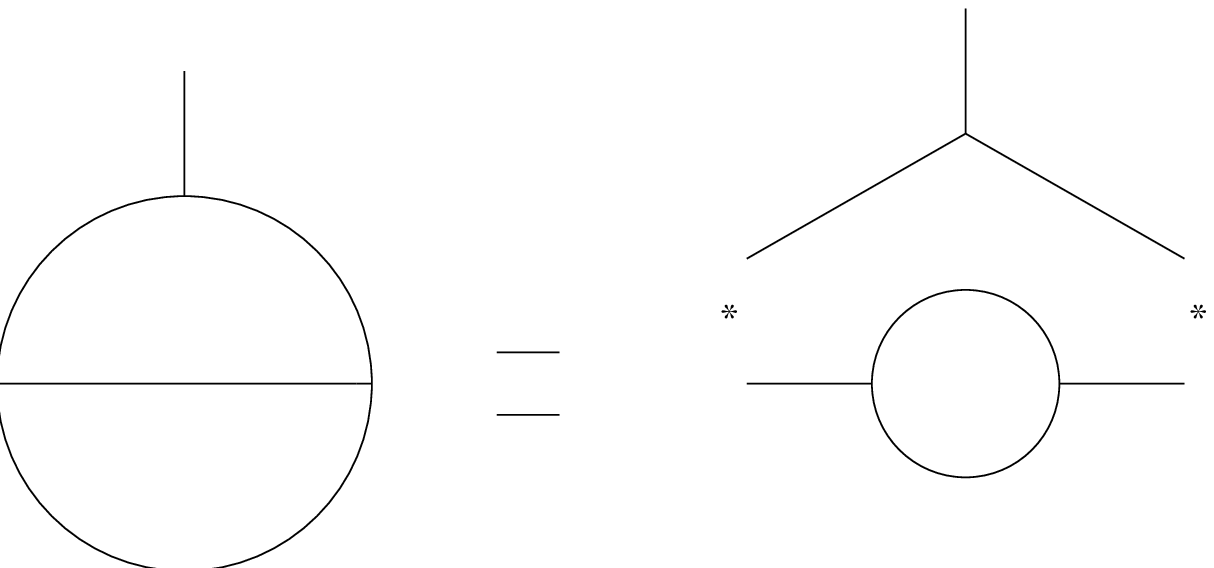,height=3cm,width=7cm}}

{\bf Fig.~2:} {\ninepoint Representation of how a two loop tadpole comes from 
contracting the one loop self-energy, 
\vskip -8pt \noindent \hglue 2.40truecm through propagators, into the outer
vertex.}

\vskip 0.5cm

In Section II of this paper we define the self-energy, explain how it was 
calculated, and give our result for it. In Section III we define the flat space
limit, compute it for the one loop graviton self-energy, and show that it 
agrees with the flat space result obtained earlier by Capper.$^7$ In Section IV
we derive the Ward identity for our gauge and describe the procedure used to 
check it. Our conclusions are discussed in Section V.

\vskip .5cm
\centerline{II. THE ONE LOOP SELF-ENERGY}

The self-energy of a quantum field is usually defined in momentum space. This 
is not convenient for our problem because the curved background prevents the 
free theory from being diagonal in a fourier basis. However, it is simple to
translate the usual prescription into a position space version which we can use.
Consider an uncharged scalar field $\phi(x)$ which has physical mass $m$, field
strength $Z$, and zero vacuum expectation value. In a flat, spacelike metric 
background we would write the full propagator as follows:
$$\Bigl\langle {\rm out} \Bigl\vert \; T\Bigl( {\widetilde \phi}(p)
\; {\widetilde \phi}(k) \Bigr) \; \Bigr\vert {\rm in}
\Bigr\rangle = {-i Z \; (2 \pi)^4 \delta^4(p+k) \over p^2 + m^2 +
\Sigma(p^2) - i \epsilon} \eqno(2.1)$$
where ${\widetilde \phi}(p)$ is the fourier transform:
$${\widetilde \phi}(p) \equiv \int d^4x \; e^{i p \cdot x} \; 
\phi(x) \eqno(2.2)$$
This means that the effective action is:
$$\eqalignno{\Gamma[\phi] &= - {1 \over 2 Z} \int {d^4p \over (2 \pi)^4} 
\; {\widetilde \phi}^*(p) \Bigl[p^2 + m^2 + \Sigma(p^2)\Bigr] 
{\widetilde \phi}(p) + {\cal O}(\phi^3) &(2.3a) \cr
&= {1 \over 2 Z} \int d^4x \; \phi(x) \Bigl[\square - m^2\Bigr] 
\phi(x) \cr
&\qquad - {1 \over 2 Z} \int d^4x_1 d^4x_2 \; \phi(x_1) \; 
\Sigma(x_1;x_2) \; \phi(x_2) + {\cal O}(\phi^3) \qquad \qquad &(2.3b) 
\cr}$$
where the position space self-energy is:
$$\Sigma(x_1;x_2) = \int {d^4p \over (2 \pi)^4} \; e^{i p \cdot (x_2 -
x_1)} \; \Sigma(p^2) \eqno(2.4)$$

It is instructive to give the one loop expansion of the self-energy for a 
general scalar field whose classical action is $S[\phi]$. We define the 
position space $n$-point vertex as:
$${\cal V}_n(x_1,\dots,x_n) \equiv {\delta^n S[\phi] \over \delta \phi(x_1)
\dots \delta \phi(x_n)} \; \Bigl\vert_{\phi = 0} \eqno(2.5)$$
In a local theory these vertices consist of a finite number of the various
derivatives times a product of delta functions:
$${\cal V}_n(x_1,\dots,x_n) = V_n(x_1;\partial_1,\dots,\partial_n) \; 
\delta^4(x_1 - x_2) \cdots \delta^4(x_1 - x_n) \eqno(2.6)$$
Note that we allow the vertex operator $V_n$ to depend upon position. Figure 3 
gives the diagrams that contribute at one loop. In our notation the result is:
$$\eqalignno{-i \Sigma(x_1';x_1'') = -\frac12 &\int d^4x_2' d^4x_3' \; 
{\cal V}_3(x_1',x_2',x_3') \cr
& \times \int d^4x_2'' d^4x_3'' \; i\Delta(x_2';x_2'') \;
i\Delta(x_3';x_3'') \; {\cal V}_3(x_1'',x_2'',x_3'') \cr
& \quad + \frac{i}2 \int d^4x_2' \int d^4x_2'' \; {\cal V}_4(x_1',x_2',
x_1'',x_2'') \; i\Delta(x_2';x_2'') + \dots &(2.7) \cr}$$
For a local theory we can do the integrations to obtain the following form:
$$\eqalignno{\Sigma(x';x'') = -\frac{i}2 &V_3(x';\partial_1',\partial_2',
\partial_3') \; i\Delta_2(x';x'') \; i\Delta_3(x';x'')
V_3(x'';\partial_1'',\partial_2'',\partial_3'') \cr
&- \frac12 V_4(x';\partial_1',\partial_2',\partial_1'',\partial_2'') \;
i\Delta_2(x';x'') \; \delta^4(x'-x'') &(2.8) \cr}$$
The $\partial_1'$ and $\partial_1''$ derivatives act outward. The other 
derivatives act on the propagator whose subscript matches their own; for 
example, $\partial_3'$ acts on the first argument of $i\Delta_3(x';x'')$. It is
sometimes convenient to partially integrate the outer derivatives, 
$\partial_1'$ and $\partial_1''$. In this case they go to minus themselves and 
they act on all $x'$'s or $x''$'s, respectively, in the expression.

\vskip 1.3cm

\centerline{\psfig{figure=f3.eps,height=3.1cm,width=13.1cm}}

{\bf Fig.~3:} {\ninepoint One loop contributions to the scalar self-energy.}

\vskip 0.5cm

It is remarkable that at one loop the position space self-energy involves no
integrations. This is why it exists at all in the in-out formalism for quantum
general relativity in a de Sitter background. The volume factors in the
interaction vertices of this theory grow so rapidly that in-out matrix elements
are generally infrared divergent if they contain even a single integration.$^{
1,8}$ Of course the higher loop contributions do contain such integrations, so 
we cannot speak of an in-out self-energy beyond one loop. A related point is 
that only the first term of (2.8) is non-zero for $x' \neq x''$. The infrared
properties of the one loop self-energy are entirely controlled by the first
term --- Fig.~3b --- and it is only the analog of this first term that we shall
study in quantum general relativity. Note as well that the first term of (2.8) 
is completely well defined for $x'$ and $x''$ away from coincidence.

The invariant Lagrangian of general relativity is:
$${\cal L} = {1 \over 16 \pi G} \Bigl(R - 2 \Lambda\Bigr) \sqrt{-g} + \Bigl({
\rm counterterms}\Bigr) \eqno(2.9)$$
where $G$ is Newton's constant and $\Lambda$ is the cosmological constant. Our
classical background has the homogeneous and isotropic form (1.1) with scale
factor:
$${\rm a}_{\rm class}(t) = e^{Ht} \eqno(2.10)$$
It is simplest to perform the calculation in conformally flat coordinates, 
for which the invariant element of the background is:
$$-dt^2 + {\rm a}^2_{\rm class}(t) \; d{\vec x} \cdot d{\vec x} = \Omega^2
\Bigl(-du^2 + d{\vec x} \cdot d{\vec x}\Bigr) \eqno(2.11a)$$
$$\Omega \equiv {1 \over H u} = \exp(H t) \eqno(2.11b)$$
Note the temporal inversion and the fact that the onset of inflation at 
$t=0$ corresponds to $u = H^{-1}$. Since the infinite future is at $u = 
0^+$, and since the spatial coordinates fall within the region, $-\frac12 
H^{-1} < x^i \leq \frac12 H^{-1}$, the range of conformal coordinates is 
rather small. This is why a conformally invariant field --- whose dynamics 
are locally the same as in flat space, except for ultraviolet regularization 
--- cannot induce a big infrared effect.

Perturbation theory is organized most conveniently in terms of a 
``pseudo-graviton'' field, $\psi_{\mu \nu}$, obtained by conformally 
re-scaling the metric:
$$g_{\mu \nu} \equiv \Omega^2 \; {\widetilde g}_{\mu \nu} \equiv 
\Omega^2 \; \Bigl(\eta_{\mu \nu} + \kappa \psi_{\mu \nu}\Bigr) 
\eqno(2.12)$$
As usual, pseudo-graviton indices are raised and lowered with the Lorentz
metric,\footnote{*}{\tenpoint Note, however, that ${\widetilde g}^{\mu\nu}$ is 
the full matrix inverse of ${\widetilde g}_{\mu\nu}$ and has the usual 
geometric series expansion:
$${\widetilde g}_{\mu\nu} = \eta^{\mu\nu} - \kappa \psi^{\mu\nu} + \kappa^2
\psi^{\mu\rho} \psi_{\rho}^{~\nu} - \dots$$}
and the loop counting parameter is $\kappa^2 \equiv 16 \pi G$. After some 
judicious partial integrations the invariant part of the bare Lagrangian takes 
the following form:$^9$
$$\eqalignno{{\cal L}_{\rm inv} =
\sqrt{-{\widetilde g}} \; {\widetilde g}^{\alpha \beta} \;
{\widetilde g}^{\rho \sigma} \; {\widetilde g}^{\mu \nu} 
\Bigl[\half \psi_{\alpha \rho , \mu} \; \psi_{\nu \sigma , \beta} - 
\half \psi_{\alpha \beta , \rho} &\; \psi_{\sigma \mu , \nu} + 
\frac14 \psi_{\alpha \beta , \rho} \; \psi_{\mu \nu , \sigma} - 
\frac14 \psi_{\alpha \rho , \mu} \; \psi_{\beta \sigma , \nu}\Bigr] 
\Omega^{2} \cr &- \half \sqrt{-{\widetilde g}} \; 
{\widetilde g}^{\rho \sigma} \; {\widetilde g}^{\mu \nu} \; 
\psi_{\rho \sigma , \mu} \; \psi_{\nu}^{~\alpha} \; 
(\Omega^{2})_{,\alpha} &(2.13) \cr}$$
Note that each interaction term contains at least one ordinary derivative. 
This occurs because the dimension three coupling is canceled by the 
undifferentiated terms from the covariant derivatives of the dimension 
five coupling. Such a cancellation --- for which there is no scalar field 
or flat space analog --- is essential for classical stability$^{10}$ against 
growth of zero modes. An interesting consequence is that the leading 
infrared effects cancel as well in the quantum theory. However, the two 
couplings do not agree at subleading order, and there is still a very 
strong quantum effect.

Gauge fixing is accomplished through the addition of $-\half \eta^{\mu \nu} 
F_{\mu} F_{\nu}$ where:$^9$
$$F_{\mu} \equiv \Bigl(\psi^{\rho}_{~\mu , \rho} - 
\frac12 \psi^{\rho}_{~\rho , \mu} + 2 \psi^{\rho}_{~\mu} \; 
{(\ln \Omega)}_{,\rho}\Bigr) \Omega \eqno(2.14)$$
The associated ghost Lagrangian is:$^9$
$$\eqalignno{{\cal L}_{\rm ghost} = -\Omega^2 \; 
&{\overline \omega}^{\mu , \nu} \; \Bigl[{\widetilde g}_{\rho \mu} \; 
\partial_{\nu} + {\widetilde g}_{\rho \nu} \; \partial_{\mu} + 
{\widetilde g}_{\mu \nu , \rho} + 2 {\widetilde g}_{\mu \nu} \; 
{(\ln \Omega)}_{, \rho} \Bigr] \; \omega^{\rho} \cr 
&+ {\Bigl( \Omega^2 \; {\overline \omega}^{\mu} \Bigr)}_{, \mu} 
\eta^{\rho \sigma} \; \Bigl[{\widetilde g}_{\nu \rho} \; 
\partial_{\sigma} + \frac12 {\widetilde g}_{\rho \sigma , \nu} + 
{\widetilde g}_{\rho \sigma} \; {(\ln \Omega)}_{, \nu} \Bigr] \; 
\omega^{\nu} &(2.15) \cr}$$
The zeroth order action results in the following free field expansion:$^{11}$
$$\psi_{\mu \nu}(u,{\vec x}) = 
\Biggl({{\rm Zero} \atop {\rm Modes}}\Biggr) + 
H^3 \sum_{\lambda, {\vec k}\neq 0} \Biggl\{ \Psi_{\mu \nu}\Bigl(u,{\vec x};
{\vec k},\lambda\Bigr) \; a({\vec k},\lambda) + \Psi_{\mu \nu}^*
\Bigl(u,{\vec x};{\vec k},\lambda\Bigr) \; a^{\dagger}({\vec k},\lambda) 
\Biggr\} \eqno(2.16)$$
The spatial polarizations consist of ``A'' modes:
$$\Psi_{\mu \nu}\Bigl(u,{\vec x};{\vec k},\lambda\Bigr) = 
{Hu \over \sqrt{2 k}} \; \Bigl(1 + {i \over k u}\Bigr) \; 
\exp\Bigl[i k \Bigl(u - \frac1{H}\Bigr) + 
i {\vec k} \cdot {\vec x}\Bigr] \; \epsilon_{\mu \nu}({\vec k},\lambda) 
\qquad \forall \lambda \in A \eqno(2.17a)$$
while the space--time and purely temporal polarizations are associated, 
respectively, with ``B'' and ``C'' modes:
$$\Psi_{\mu \nu}\Bigl(u,{\vec x};{\vec k},\lambda\Bigr) = 
{Hu \over \sqrt{2 k}} \; \exp\Bigl[i k \Bigl(u - \frac1{H}\Bigr) + 
i {\vec k} \cdot {\vec x} \Bigr] \; \epsilon_{\mu \nu}({\vec k},\lambda) 
\qquad \forall \lambda \in B,C \eqno(2.17b)$$
In LSZ reduction one would integrate against and contract into $\Psi_{\mu 
\nu}(u,{\vec x};{\vec k},\lambda)$ to insert and ``in''-coming graviton of 
momentum ${\vec k}$ and polarization $\lambda$; the conjugate would be used 
to extract an ``out''-going graviton with the same quantum numbers. The 
zero modes evolve as free particles with time dependences $1$ and $u^3$ 
for the A modes, and $u$ and $u^2$ for the B and C modes. Since causality 
decouples the zero modes shortly after the onset of inflation, they play 
no role in screening and we shall not trouble with them further.

We define $\vert 0 \rangle$ as the Heisenberg state annihilated by 
$a({\vec k}, \lambda)$ --- and the analogous ghost operators --- at 
the onset of inflation. We can use this condition and expansion (2.16) 
to express the free pseudo-graviton propagator as a mode sum:$^8$
$$\eqalignno{i\Bigl[ {_{\mu \nu}}\Delta_{\rho \sigma}\Bigr](x;x') &\equiv
\Bigl\langle 0 \Bigl\vert T\Bigl\{\psi_{\mu \nu}(x) \; 
\psi_{\rho \sigma}(x')\Bigr\} \Bigr\vert 0 \Bigr\rangle_{\rm free} 
&(2.18a) \cr 
&= H^3 \sum_{\lambda, {\vec k} \neq 0} \Biggl\{ \theta(u'-u) \; \Psi_{\mu \nu}
\; {\Psi'}^*_{\rho \sigma} + \theta(u-u') \; \Psi^*_{\mu \nu} \; {\Psi'}_{\rho 
\sigma} \Biggr\} e^{- \epsilon \Vert {\vec k} \Vert} \qquad &(2.18b) \cr}$$
Note that the convergence factor $e^{- \epsilon \Vert {\vec k} \Vert}$ serves 
as an ultraviolet mode cutoff. Although the resulting regularization is very 
convenient for this calculation, its failure to respect general coordinate 
invariance necessitates the use of non-invariant counterterms. These are 
analogous to the photon mass which must be added to QED when using a momentum 
cutoff. Just as in QED, these non-invariant counterterms do not affect long 
distance phenomena.

Because the propagator is only needed for small conformal coordinate 
separations, ${\Delta x} \equiv \Vert {\vec x}' - {\vec x} \Vert$ and 
${\Delta u} \equiv u' - u$, the sum over momenta is well approximated as 
an integral. When this is done the pseudo-graviton and ghost propagators 
become:$^8$
$$\eqalignno{i \Bigl[{_{\mu \nu}} \Delta_{\rho \sigma}\Bigr](x;x') &\approx 
\; {H^2 \over 8 {\pi}^2} \; \Biggl\{ {2u'u \over {\Delta x}^2 - 
{\Delta u}^2 + 2 i \epsilon \vert {\Delta u} \vert + \epsilon^2} \;
\Bigl[{2 \eta_{\mu (\rho} \; \eta_{\sigma) \nu}} - \eta_{\mu \nu} \; 
\eta_{\rho \sigma} \Bigr] \cr
&- \ln \Bigl[H^2 \Bigl({\Delta x}^2 - {\Delta u}^2 + 
2 i \epsilon \vert {\Delta u} \vert + \epsilon^2\Bigr) \Bigr] \;
\Bigl[ 2{\overline \eta}_{\mu(\rho} \;
{\overline \eta}_{\sigma)\nu} - 2 {\overline \eta_{\mu \nu}} \;
{\overline \eta_{\rho \sigma}} \Bigr] \; \Biggr\} \qquad \qquad &(2.19a) \cr}$$
$$\eqalignno{i \Bigl[{_{\mu}} \Delta_{\nu}\Bigr](x;x') \approx 
\; {H^2 \over 8 {\pi}^2} \; \Biggl\{ &{2u'u \over {\Delta x}^2 - 
{\Delta u}^2 + 2 i \epsilon \vert {\Delta u} \vert + \epsilon^2} \;
\eta_{\mu\nu}\cr 
&\qquad - \ln\Bigl[H^2 \Bigl({\Delta x}^2 - {\Delta u}^2 + 
2 i \epsilon \vert {\Delta u} \vert + \epsilon^2\Bigr)\Bigr] \;
{\overline \eta_{\mu \nu}} \Biggr\} &(2.19b) \cr}$$
Parenthesized indices are symmetrized and a bar above a Lorentz metric 
or a Kronecker delta symbol means that the zero component is projected 
out, e.g. ${\overline \eta}_{\mu \nu} \equiv \eta_{\mu \nu} + \delta_
{\mu}^{~0} \; \delta_{\nu}^{~0}$. The decoupling between functional 
dependence upon spacetime and tensor indices --- and the simplicity of 
each --- greatly facilitates calculations.

We find the cubic self-interactions by expanding expression (2.13) to third 
order in the pseudo-graviton field. The result is $\kappa \Omega^2$ times:
$$-\frac1{2u} \psi \; \psi_{,\mu} \; \psi^{\mu\nu} \; 
t_{\nu} + \frac1{u} \psi^{\rho\sigma} \; \psi_{\rho\sigma,\mu} 
\; \psi^{\mu\nu} \; t_{\nu} + \frac1{u} \psi_{\rho\sigma}
\; \psi^{,\rho} \; \psi^{\sigma\nu} \; t_{\nu}$$
$$\frac14 \psi \; \psi^{\rho\sigma,\mu} \; \psi_{\mu\rho,
\sigma} - \psi^{\rho\sigma} \; \psi_{\rho}^{~\mu,\nu} \; 
\psi_{\mu\nu,\sigma} -\frac12 \psi^{\rho\sigma} \; \psi_{\rho}^{~\mu,
\nu} \; \psi_{\sigma\nu,\mu}$$
$$-\frac14 \psi \; \psi_{,\rho} \; \psi^{\rho\sigma}_{~~,\sigma}
+ \frac12 \psi^{\rho\sigma} \; \psi_{\rho\sigma,\mu} \; \psi^{
\mu\nu}_{~~,\nu} + \frac12 \psi^{\rho\sigma} \; \psi_{,\rho} \;
\psi_{\sigma\mu}^{~~~,\mu} + \frac12 \psi^{\rho\sigma} \; \psi^{,\mu}
\; \psi_{\mu\rho,\sigma} \eqno(2.20)$$
$$+\frac18 \psi \; \psi^{,\mu} \; \psi_{,\mu} - \frac12 \psi^{
\rho\sigma} \; \psi_{\rho\sigma,\mu} \; \psi^{,\mu} -\frac14 
\psi^{\rho\sigma} \; \psi_{,\rho} \; \psi_{,\sigma}$$
$$-\frac18 \psi \; \psi^{\rho\sigma,\mu} \; \psi_{\rho\sigma,
\mu} + \frac12 \psi^{\rho\sigma} \; \psi_{\rho\mu,\nu} \; 
\psi_{\sigma}^{~\mu,\nu} + \frac14 \psi^{\rho\sigma} \; \psi_{\mu\nu,
\rho} \; \psi^{\mu\nu}_{~~~,\sigma}$$
where $\psi \equiv \psi^{\mu}_{~\mu}$ and $t_{\nu} \equiv \eta_{0\nu}$. All but
the three terms of the first line should agree with the flat space expansion 
when $\psi_{\mu\nu}$ is regarded as the graviton field and $\Omega = 1$. Of 
course this allows us to check them against published results,$^{12}$ and they 
do check.\footnote{*}{\tenpoint Note, however, that the earlier results are 
given for a timelike metric, so our field is minus theirs.}

The vertex operators  will be fully symmetrized if we define them by functional 
differentiation as in (2.5). This is not efficient for computing the 
self-energy because only one of the vertices needs to be symmetrized on its
internal lines. Fully symmetrizing both vertex operators causes each distinct 
pairing to appear twice in the self-energy, which is why the symmetry factor 
for this diagram is $\frac12$. The large number of distinct cubic 
self-interaction terms (2.20) means that this is not an efficient strategy for 
quantum gravity. The fully symmetrized vertex operator contains 75 separate 
terms,$^4$ whereas permuting over only the three possible choices for the outer
line results in just 43 terms. One must sum over both ways of pairing the 
internal lines but there is still a saving of almost 50\% --- which is 
important in summing over $75^2$ eight-fold contractions of 20--index objects!

To obtain the partially symmetrized vertex operators from a given cubic 
self-interaction one merely assigns the three legs any of the six possible ways
and then permutes cyclicly. For example, the first term in (2.20) gives:
$$\eqalignno{-\frac1{2u} \psi \; \psi_{\mu} \; \psi^{\mu \nu}
\; t_{\nu} &\longrightarrow -\frac1{2u} \eta^{\alpha_1 \beta_1} 
\; \eta^{\alpha_2 \beta_2} \; \partial_2^{(\alpha_3} \;
t^{\beta_3)} &({\rm Initial\ Assignment}) \cr
&\longrightarrow -\frac1{2u} \eta^{\alpha_1 \beta_1} \; \eta^{\alpha_2 
\beta_2} \; \partial_2^{(\alpha_3} \; t^{\beta_3)} -\frac1{2u} 
\eta^{\alpha_2 \beta_2} \; \eta^{\alpha_3 \beta_3} \; 
\partial_3^{(\alpha_1} \; t^{\beta_1)} \cr
&\qquad -\frac1{2u} \eta^{\alpha_3 \beta_3} \; \eta^{\alpha_1 
\beta_1} \; \partial_1^{(\alpha_2} \; t^{\beta_2)} &({\rm 
Cyclic\ Permutation}) \cr}$$
The various partial vertex operators are given in Table~1. We consider line \#1
to be the distinguished one, and we have used symmetries among the remaining 
two fields to reduce the number of vertices whenever possible. 

The ghost-anti-ghost-pseudo-graviton interactions can be read off from (2.15).
There are only ten interactions and they are $\kappa \Omega^2$ times:
$$-\psi_{\mu \nu} \; {\overline \omega}^{\mu,\rho} \; \omega^{
\nu}_{~,\rho} - \psi_{\mu\nu} \; {\overline \omega}^{\rho,\mu} 
\; \omega^{\nu}_{~,\rho} - \psi_{\mu\nu,\sigma} \; {\overline
\omega}^{\mu,\nu} \; \omega^{\sigma} - \frac2{u} \psi_{\mu\nu} 
\; {\overline \omega}^{\mu,\nu} \; \omega^{\sigma} \; 
t_{\sigma} + \psi_{\mu\nu} \; {\overline \omega}^{\rho}_{~,\rho} 
\; \omega^{\mu,\nu}$$
$$+ \frac12 \psi_{,\sigma} \; {\overline \omega}^{\rho}_{~,\rho}
\; \omega^{\sigma} + \frac1{u} \psi \; {\overline \omega}^{\rho
}_{~,\rho} \; \omega^{\sigma} \; t_{\sigma} + \frac2{u} \psi_{
\mu\nu} \; {\overline \omega}^{\rho} \; \omega^{\mu,\nu}
\; t_{\rho} + \frac1{u} \psi_{,\sigma} \; {\overline \omega}^{
\rho} \; \omega^{\sigma} \; t_{\rho} + \frac2{u^2} \psi
\; {\overline \omega}^{\rho} \; \omega^{\sigma} \; 
t_{\rho} \; t_{\sigma} \eqno(2.21)$$
There is no issue of symmetrization in finding the associated vertex operators
because each of the three fields is different. The result is presented in 
Table~2.

\vfill\eject

\vbox{\tabskip=0pt \offinterlineskip
\def\tablerule{\noalign{\hrule}}
\halign to450pt {\strut#& \vrule#\tabskip=1em plus2em& \hfil#& \vrule#& 
\hfil#\hfil& \vrule#& \hfil#& \vrule#& \hfil#\hfil& \vrule#\tabskip=0pt\cr
\tablerule
\omit & height2pt & \omit && \omit && \omit && \omit &\cr
&&\omit\hidewidth \# &&\omit\hidewidth {\rm Partial Vertex}\hidewidth&& 
\omit\hidewidth \#\hidewidth&& \omit\hidewidth {\rm Partial Vertex}
\hidewidth&\cr
\omit & height2pt & \omit && \omit && \omit && \omit &\cr
\tablerule
\omit & height2pt & \omit && \omit && \omit && \omit &\cr
&& 1 && $-\frac1{2u} \eta^{\alpha_1 \beta_1} \; \eta^{\alpha_2 \beta_2}
\; \partial_2^{(\alpha_3} \; t^{\beta_3)}$ && 22 && $\frac12 
\eta^{\alpha_1 (\alpha_2} \; \eta^{\beta_2) \beta_1} \; 
\partial_2^{(\alpha_3} \; \partial_3^{\beta_3)}$ &\cr
\omit & height2pt & \omit && \omit && \omit && \omit &\cr
\tablerule
\omit & height2pt & \omit && \omit && \omit && \omit &\cr
&& 2 && $-\frac1{2u} \eta^{\alpha_2 \beta_2} \; \eta^{\alpha_3 \beta_3}
\; \partial_3^{(\alpha_1} \; t^{\beta_1)}$ && 23 && $\frac12 
\eta^{\alpha_2 (\alpha_3} \; \eta^{\beta_3) \beta_2} \;
\partial_3^{(\alpha_1} \; \partial_1^{\beta_1)}$ &\cr
\omit & height2pt & \omit && \omit && \omit && \omit &\cr
\tablerule
\omit & height2pt & \omit && \omit && \omit && \omit &\cr
&& 3 && $-\frac1{2u} \eta^{\alpha_3 \beta_3} \; \eta^{\alpha_1 \beta_1}
\; \partial_1^{(\alpha_2} \; t^{\beta_2)}$ && 24 && $\frac12 
\eta^{\alpha_3 (\alpha_1} \; \eta^{\beta_1) \beta_3} \;
\partial_1^{(\alpha_2} \; \partial_2^{\beta_2)}$ &\cr
\omit & height2pt & \omit && \omit && \omit && \omit &\cr
\tablerule
\omit & height2pt & \omit && \omit && \omit && \omit &\cr
&& 4 && $\frac1{u} \eta^{\alpha_1 (\alpha_2} \; \eta^{\beta_2) \beta_1}
\; \partial_2^{(\alpha_3} \; t^{\beta_3)}$ && 25 && $\frac12 
\partial_2^{(\alpha_1} \; \eta^{\beta_1) (\alpha_3} \; 
\partial_3^{\beta_3)} \; \eta^{\alpha_2 \beta_2}$ &\cr
\omit & height2pt & \omit && \omit && \omit && \omit &\cr
\tablerule
\omit & height2pt & \omit && \omit && \omit && \omit &\cr
&& 5 && $\frac1{u} \eta^{\alpha_2 (\alpha_3} \; \eta^{\beta_3) \beta_2}
\; \partial_3^{(\alpha_1} \; t^{\beta_1)}$ && 26 && $\frac12 
\partial_3^{(\alpha_2} \; \eta^{\beta_2) (\alpha_1} \; 
\partial_1^{\beta_1)} \; \eta^{\alpha_3 \beta_3}$ &\cr
\omit & height2pt & \omit && \omit && \omit && \omit &\cr
\tablerule
\omit & height2pt & \omit && \omit && \omit && \omit &\cr
&& 6 && $\frac1{u} \eta^{\alpha_3 (\alpha_1} \; \eta^{\beta_1) \beta_3}
\; \partial_1^{(\alpha_2} \; t^{\beta_2)}$ && 27 && $\frac12 
\partial_1^{(\alpha_3} \; \eta^{\beta_3) (\alpha_2} \; 
\partial_2^{\beta_2)} \; \eta^{\alpha_1 \beta_1}$ &\cr
\omit & height2pt & \omit && \omit && \omit && \omit &\cr
\tablerule
\omit & height2pt & \omit && \omit && \omit && \omit &\cr
&& 7 && $\frac1{u} t^{(\alpha_3} \; \eta^{\beta_3) (\alpha_1}
\; \partial_2^{\beta_1)} \; \eta^{\alpha_2 \beta_2}$ && 28 && 
$\frac12 \partial_2^{(\alpha_1} \; \eta^{\beta_1) (\alpha_2} \;
\partial_3^{\beta_2)} \; \eta^{\alpha_3 \beta_3}$ &\cr
\omit & height2pt & \omit && \omit && \omit && \omit &\cr
\tablerule
\omit & height2pt & \omit && \omit && \omit && \omit &\cr
&& 8 && $\frac1{u} t^{(\alpha_1} \; \eta^{\beta_1) (\alpha_2} 
\; \partial_3^{\beta_2)} \; \eta^{\alpha_3 \beta_3}$ && 29 && 
$\frac12 \partial_3^{(\alpha_2} \; \eta^{\beta_2) (\alpha_3} \; 
\partial_1^{\beta_3)} \; \eta^{\alpha_1 \beta_1}$ &\cr
\omit & height2pt & \omit && \omit && \omit && \omit &\cr
\tablerule
\omit & height2pt & \omit && \omit && \omit && \omit &\cr
&& 9 && $\frac1{u} t^{(\alpha_2} \; \eta^{\beta_2) (\alpha_3} 
\; \partial_1^{\beta_3)} \; \eta^{\alpha_1 \beta_1}$ && 30 && 
$\frac12 \partial_1^{(\alpha_3} \; \eta^{\beta_3) (\alpha_1} \; 
\partial_2^{\beta_1)} \; \eta^{\alpha_2 \beta_2}$ &\cr
\omit & height2pt & \omit && \omit && \omit && \omit &\cr
\tablerule
\omit & height2pt & \omit && \omit && \omit && \omit &\cr
&& 10 && $\frac14 \eta^{\alpha_1 \beta_1} \; \partial_3^{(\alpha_2} 
\; \eta^{\beta_2) (\alpha_3} \; \partial_2^{\beta_3)}$ && 31 && 
$\frac18 \eta^{\alpha_1 \beta_1} \; \eta^{\alpha_2 \beta_2} \;
\eta^{\alpha_3 \beta_3} \; \partial_2 \cdot \partial_3$ &\cr
\omit & height2pt & \omit && \omit && \omit && \omit &\cr
\tablerule
\omit & height2pt & \omit && \omit && \omit && \omit &\cr
&& 11 && $\frac14 \eta^{\alpha_2 \beta_2} \; \partial_1^{(\alpha_3} 
\; \eta^{\beta_3) (\alpha_1} \; \partial_3^{\beta_1)}$ && 32 && 
$\frac14 \eta^{\alpha_1 \beta_1} \; \eta^{\alpha_2 \beta_2} \;
\eta^{\alpha_3 \beta_3} \; \partial_3 \cdot \partial_1$ &\cr
\omit & height2pt & \omit && \omit && \omit && \omit &\cr
\tablerule
\omit & height2pt & \omit && \omit && \omit && \omit &\cr
&& 12 && $\frac14 \eta^{\alpha_3 \beta_3} \; \partial_2^{(\alpha_1} 
\; \eta^{\beta_1) (\alpha_2} \; \partial_1^{\beta_2)}$ && 33 && 
$-\frac12 \eta^{\alpha_1 (\alpha_2} \; \eta^{\beta_2) \beta_1} 
\; \eta^{\alpha_3 \beta_3} \; \partial_2 \cdot \partial_3$ &\cr
\omit & height2pt & \omit && \omit && \omit && \omit &\cr
\tablerule
\omit & height2pt & \omit && \omit && \omit && \omit &\cr
&& 13 && $-\partial_3^{(\alpha_1} \; \eta^{\beta_1) (\alpha_2} 
\; \eta^{\beta_2) (\alpha_3} \; \partial_2^{\beta_3)}$ && 34 && 
$-\frac12 \eta^{\alpha_2 (\alpha_3} \; \eta^{\beta_3) \beta_2}
\; \eta^{\alpha_1 \beta_1} \; \partial_3 \cdot \partial_1$ &\cr
\omit & height2pt & \omit && \omit && \omit && \omit &\cr
\tablerule
\omit & height2pt & \omit && \omit && \omit && \omit &\cr
&& 14 && $-\partial_1^{(\alpha_2} \; \eta^{\beta_2) (\alpha_3} 
\; \eta^{\beta_3) (\alpha_1} \; \partial_3^{\beta_1)}$ && 35 &&
$-\frac12 \eta^{\alpha_3 (\alpha_1} \; \eta^{\beta_1) \beta_3}
\; \eta^{\alpha_2 \beta_2} \; \partial_1 \cdot \partial_2$ &\cr
\omit & height2pt & \omit && \omit && \omit && \omit &\cr
\tablerule
\omit & height2pt & \omit && \omit && \omit && \omit &\cr
&& 15 && $-\partial_2^{(\alpha_3} \; \eta^{\beta_3) (\alpha_1} 
\; \eta^{\beta_1) (\alpha_2} \; \partial_1^{\beta_2)}$ && 36 && 
$-\frac14 \partial_2^{(\alpha_1} \; \partial_3^{\beta_1)} \;
\eta^{\alpha_2 \beta_2} \; \eta^{\alpha_3 \beta_3}$ &\cr
\omit & height2pt & \omit && \omit && \omit && \omit &\cr
\tablerule
\omit & height2pt & \omit && \omit && \omit && \omit &\cr
&& 16 && $-\frac12 \partial_3^{(\alpha_2} \; \eta^{\beta_2) (\alpha_1} 
\; \eta^{\beta_1) (\alpha_3} \; \partial_2^{\beta_3)}$ && 37 &&
$-\frac12 \partial_3^{(\alpha_2} \; \partial_1^{\beta_2)} \;
\eta^{\alpha_3 \beta_3} \; \eta^{\alpha_1 \beta_1}$ &\cr
\omit & height2pt & \omit && \omit && \omit && \omit &\cr
\tablerule
\omit & height2pt & \omit && \omit && \omit && \omit &\cr
&& 17 && $-\frac12 \partial_1^{(\alpha_3} \; \eta^{\beta_3) (\alpha_2} 
\; \eta^{\beta_2) (\alpha_1} \; \partial_3^{\beta_1)}$ && 38 &&
$-\frac18 \eta^{\alpha_1 \beta_1} \; \eta^{\alpha_2 (\alpha_3} 
\; \eta^{\beta_3) \beta_2} \; \partial_2 \cdot \partial_3$ &\cr
\omit & height2pt & \omit && \omit && \omit && \omit &\cr
\tablerule
\omit & height2pt & \omit && \omit && \omit && \omit &\cr
&& 18 && $-\frac12 \partial_2^{(\alpha_1} \; \eta^{\beta_1) (\alpha_3} 
\; \eta^{\beta_3) (\alpha_2} \; \partial_1^{\beta_2)}$ && 39 &&
$-\frac14 \eta^{\alpha_2 \beta_2} \; \eta^{\alpha_3 (\alpha_1} 
\; \eta^{\beta_1) \beta_3} \; \partial_3 \cdot \partial_1$ &\cr
\omit & height2pt & \omit && \omit && \omit && \omit &\cr
\tablerule
\omit & height2pt & \omit && \omit && \omit && \omit &\cr
&& 19 && $-\frac14 \eta^{\alpha_1 \beta_1} \; \eta^{\alpha_2 \beta_2}
\; \partial_2^{(\alpha_3} \; \partial_3^{\beta_3)}$ && 40 &&
$\frac12 \eta^{\alpha_1) (\alpha_2} \; \eta^{\beta_2) (\alpha_3} 
\; \eta^{\beta_3) (\beta_1} \; \partial_2 \cdot \partial_3$ &\cr
\omit & height2pt & \omit && \omit && \omit && \omit &\cr
\tablerule
\omit & height2pt & \omit && \omit && \omit && \omit &\cr
&& 20 && $-\frac14 \eta^{\alpha_2 \beta_2} \; \eta^{\alpha_3 \beta_3} 
\; \partial_3^{(\alpha_1} \; \partial_1^{\beta_1)}$ && 41 &&
$\eta^{\alpha_1) (\alpha_2} \; \eta^{\beta_2) (\alpha_3} \; 
\eta^{\beta_3) (\beta_1} \; \partial_3 \cdot \partial_1$ &\cr
\omit & height2pt & \omit && \omit && \omit && \omit &\cr
\tablerule
\omit & height2pt & \omit && \omit && \omit && \omit &\cr
&& 21 && $-\frac14 \eta^{\alpha_3 \beta_3} \; \eta^{\alpha_1 \beta_1} 
\; \partial_1^{(\alpha_2} \; \partial_2^{\beta_2)}$ && 42 &&
$\frac14 \partial_2^{(\alpha_1} \; \partial_3^{\beta_1)} \;
\eta^{\alpha_2 (\alpha_3} \; \eta^{\beta_3) \beta_2}$ &\cr
\omit & height2pt & \omit && \omit && \omit && \omit &\cr
\tablerule
\omit & height2pt & \omit && \omit && \omit && \omit &\cr
&& \omit && \omit && 43 && $\frac12 \partial_3^{(\alpha_2} \; 
\partial_1^{\beta_2)} \; \eta^{\alpha_3 (\alpha_1} \; 
\eta^{\beta_1) \beta_3}$ &\cr 
\omit & height2pt & \omit && \omit && \omit && \omit &\cr
\tablerule}}

{\bf Table~1:} {\ninepoint Cubic partial vertex operators with \#1 
distinguished. Each term should be multiplied by $\kappa \Omega^2$.}

\vfill\eject

\vbox{\tabskip=0pt \offinterlineskip
\def\tablerule{\noalign{\hrule}}
\halign to450pt {\strut#& \vrule#\tabskip=1em plus2em& \hfil#& \vrule#& 
\hfil#\hfil& \vrule#& \hfil#& \vrule#& \hfil#\hfil& \vrule#\tabskip=0pt\cr
\tablerule
\omit & height2pt & \omit && \omit && \omit && \omit &\cr
&&\omit\hidewidth \# &&\omit\hidewidth {\rm Vertex Operator}\hidewidth&& 
\omit\hidewidth \#\hidewidth&& \omit\hidewidth {\rm Vertex Operator}
\hidewidth&\cr
\omit & height2pt & \omit && \omit && \omit && \omit &\cr
\tablerule
\omit & height2pt & \omit && \omit && \omit && \omit &\cr
&& 1 && $- \eta^{\alpha_2 (\alpha_1} \; \eta^{\beta_1) \alpha_3}
\; \partial_2 \cdot \partial_3$ && 6 && $\frac12 \eta^{\alpha_1 
\beta_1} \; \partial_2^{\alpha_2} \; \partial_1^{\alpha_3}$ &\cr
\omit & height2pt & \omit && \omit && \omit && \omit &\cr
\tablerule
\omit & height2pt & \omit && \omit && \omit && \omit &\cr
&& 2 && $- \eta^{\alpha_3 (\alpha_1} \; \partial_2^{\beta_1)} 
\; \partial_3^{\alpha_2}$ && 7 && $\frac1{u} \eta^{\alpha_1 \beta_1} 
\; \partial_2^{\alpha_2} \; t^{\alpha_3}$ &\cr
\omit & height2pt & \omit && \omit && \omit && \omit &\cr
\tablerule
\omit & height2pt & \omit && \omit && \omit && \omit &\cr
&& 3 && $- \eta^{\alpha_2 (\alpha_1} \; \partial_2^{\beta_1)} 
\; \partial_1^{\alpha_3}$ && 8 && $\frac2{u} \eta^{\alpha_3 
(\alpha_1} \; \partial_3^{\beta_1)} \; t^{\alpha_2}$ &\cr
\omit & height2pt & \omit && \omit && \omit && \omit &\cr
\tablerule
\omit & height2pt & \omit && \omit && \omit && \omit &\cr
&& 4 && $- \frac2{u} \eta^{\alpha_2 (\alpha_1} \; \partial_2^{\beta_1)}
\; t^{\alpha_3}$ && 9 && $\frac1{u} \eta^{\alpha_1 \beta_1} \; 
\partial_1^{\alpha_3} \; t^{\alpha_2}$ &\cr
\omit & height2pt & \omit && \omit && \omit && \omit &\cr
\tablerule
\omit & height2pt & \omit && \omit && \omit && \omit &\cr
&& 5 && $\eta^{\alpha_3 (\alpha_1} \; \partial_3^{\beta_1)} \; 
\partial_2^{\alpha_2}$ && 10 && $\frac2{u^2} \eta^{\alpha_1 \beta_1} \;
t^{\alpha_2} \; t^{\alpha_3}$ &\cr
\omit & height2pt & \omit && \omit && \omit && \omit &\cr
\tablerule}}

{\bf Table~2:} {\ninepoint Ghost-pseudo-graviton vertex operators. Each term 
should be multiplied by $\kappa \Omega^2$.}

\vskip 1cm

\centerline{\psfig{figure=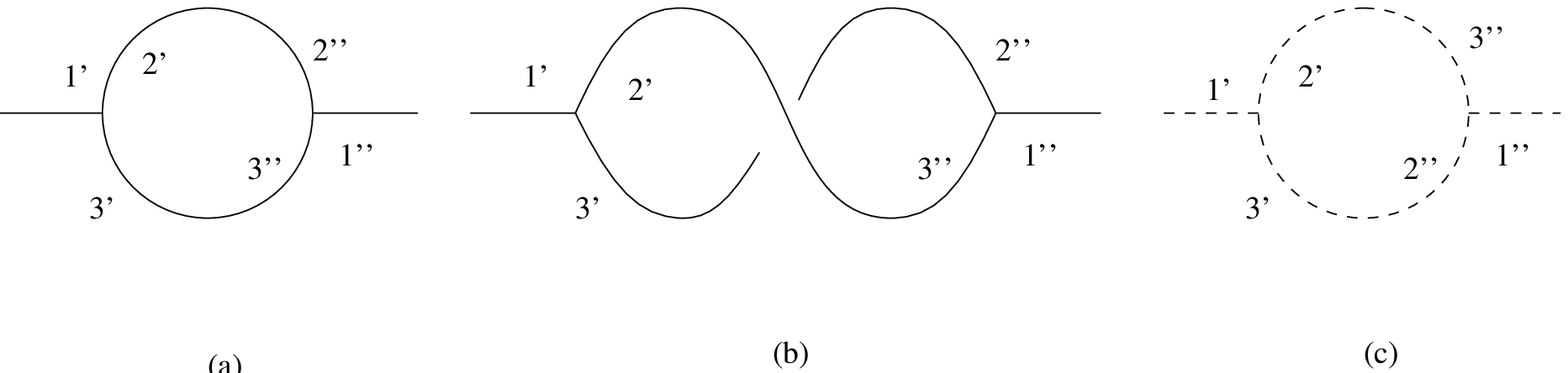,height=3.0cm,width=15.8cm}}

{\bf Fig.~4:} {\ninepoint One-loop contributions to the graviton self-energy.}

\vskip 0.5cm

It is simple to write the graviton self-energy as a sum over contractions of
the various vertex operators between the two internal propagators:
$$\eqalignno{\Bigl[{^{\alpha_1\beta_1}}&\Sigma^{\rho_1\sigma_1}\Bigr](x';x'') =
-i \sum_{i,j=1}^{43} V_i^{\alpha_1\beta_1\alpha_2\beta_2\alpha_3\beta_3}(x';
\partial_1',\partial_2',\partial_3') \enskip i\Bigl[{_{\alpha_2\beta_2}}
\Delta_{\rho_2\sigma_2} \Bigr](x';x'') \cr
&\times i\Bigl[{_{\alpha_3\beta_3}}\Delta_{\rho_3\sigma_3} \Bigr](x';x'') 
\enskip \Bigl\{V_j^{\rho_1\sigma_1\rho_2\sigma_2\rho_3\sigma_3}(x'';
\partial_1'', \partial_2'',\partial_3'') + V_j^{\rho_1\sigma_1\rho_3\sigma_3
\rho_2\sigma_2}(x'';\partial_1'',\partial_3'',\partial_2'')\Bigr\} \cr
&\qquad \qquad +i \sum_{i,j=1}^{10} V_i^{\alpha_1\beta_1\alpha_2\alpha_3}(x';
\partial_1',\partial_2',\partial_3') \enskip i\Bigl[{_{\alpha_2}} \Delta_{
\rho_2} \Bigr](x';x'') &(2.22) \cr
&\qquad \qquad \qquad \qquad \times i\Bigl[{_{\alpha_3}}\Delta_{\rho_3} 
\Bigr](x';x'') \enskip V_j^{\rho_1\sigma_1\rho_3\rho_2}(x'';\partial_1'',
\partial_3'',\partial_2'') \cr}$$
Since we have not symmetrized the pseudo-graviton vertex operators on lines 2 
and 3 it is necessary to include both a symmetric loop (Fig.~4a) and an 
asymmetric loop (Fig.~4b). The second double sum gives the ghost loop 
(Fig.~4c). The distinction between the $\psi^3$ and $\psi {\overline \omega}
\omega$ vertex operators does not require a separate symbol; the different 
number of indices suffices. Where each derivative acts is denoted by primes and
subscripts. For example, the derivative $\partial_2'$ in the first $\psi^3$ 
vertex operator acts on the first argument of the propagator, $\Bigl[{_{\alpha_2
\beta_2}}\Delta_{\rho_2\sigma_2}\Bigr](x';x'')$. The derivatives $\partial_1'$ 
and $\partial_1''$ act on the left and right outer legs. In computing the two 
loop tadpole we left them as free operators to act on the upper propagators 
when they were attached. For computing the self-energy we have of course 
partially integrated them to act on all the $x'$'s and $x''$'s in the 
expression. 

The entire calculation was performed by computer using the symbolic 
manipulation program Mathematica.$^{13}$ The first step was to contract each
pair of vertex operators into the internal propagators. This was done using 
Mertig's package FeynCalc,$^{14}$ and the result was written onto a file. The 
next step was acting the internal derivatives ($\partial_2'$, $\partial_2''$, 
$\partial_3'$, and $\partial_3''$), the results of which were also stored for
each pair of vertex operators. Selected vertex pairs were computed by hand to
check the procedure. At this stage the results for all vertex pairs were 
summed, and the total was checked for symmetry under interchanging the two
external legs.

The result we obtained for the self-energy operator is a very long sum of 
4-index objects times scalar functions. The 4-index objects are constructed 
from $\eta^{\mu\nu}$, $x^{\mu} \equiv (x'-x'')^{\mu}$, $t^{\mu} \equiv \delta^{
\mu}_{~0}$, ${\partial_1'}^{\mu}$ and ${\partial_1''}^{\mu}$. The scalar 
functions can depend upon $x^2 \equiv x^{\mu} x^{\nu} \eta_{\mu\nu}$, $u'$, 
$u''$, $x \cdot \partial_1'$, $x \cdot \partial_1''$, $t \cdot \partial_1'$, 
$t \cdot \partial_1''$ and $\partial_1' \cdot \partial_1''$. Each term can 
contain at most one factor of each of the external derivatives, either free or
contracted. It can be shown that with this requirement there are 79 distinct
4-index objects and ten possibilities for contracted derivatives.$^4$ To make
the expression more manageable, prior to attaching the outer vertex and 
propagators, we extracted the scalar coefficient of each allowed combination of
4-index object and contracted external derivative operators. It is from this
data that we later computed the self-energy by partially integrating the 
external derivatives to act back on $x'$ and $x''$ respectively. Since the data
actually contained $x'$ and $x''$ through $x$, $u'$, and $u''$, we used the
rules:
$$-\partial_{1 \mu}' = -{\partial \over \partial x^{\mu}} + t_{\mu} {\partial 
\over \partial u'} \eqno(2.23a)$$
$$-\partial_{1 \mu}'' = {\partial \over \partial x^{\mu}} + t_{\mu} {\partial 
\over \partial u''} \eqno(2.23b)$$
The result after this is done consists of a sum of 4-index tensors times scalar
functions. The 4-index tensors can depend only upon $\eta^{\mu\nu}$, $x^{\mu}$
and $t^{\mu}$, while the scalar functions depend only upon $x^2$, $u'$ and 
$u''$. The 21 possible 4-index objects are listed in Table~3. Note that we have
dispensed with the now-irrelevant line subscript 1. Note also that the 
reflection symmetry relates the coefficients of pairs 3 and 4, 5 and 6, 7 and 
8, 10 and 11, 14 and 15, 16 and 17, and 19 and 20.

\vskip 1cm
\vbox{\tabskip=0pt \offinterlineskip
\def\tablerule{\noalign{\hrule}}
\halign to450pt {\strut#& \vrule#\tabskip=1em plus2em& \hfil#& \vrule#& 
\hfil#\hfil& \vrule#& \hfil#& \vrule#& \hfil#\hfil& \vrule#& \hfil#& \vrule#& 
\hfil#\hfil& \vrule#\tabskip=0pt\cr
\tablerule
\omit & height2pt & \omit && \omit && \omit && \omit && \omit && \omit &\cr
&&\omit\hidewidth \# &&\omit\hidewidth {\rm Tensor}\hidewidth&& \omit\hidewidth
\#\hidewidth&& \omit\hidewidth {\rm Tensor}\hidewidth&& \omit\hidewidth 
\#\hidewidth&& \omit\hidewidth {\rm Tensor}\hidewidth&\cr
\omit & height2pt & \omit && \omit && \omit && \omit && \omit && \omit &\cr
\tablerule
\omit & height2pt & \omit && \omit && \omit && \omit && \omit && \omit &\cr
&& 1 && $\eta^{\alpha\beta} \; \eta^{\rho\sigma}$ && 8 && $x^{\alpha}
\; x^{\beta} \; \eta^{\rho\sigma}$ && 15 && $t^{(\alpha} 
\; x^{\beta)} \; t^{\rho} \; t^{\sigma}$ &\cr
\omit & height2pt & \omit && \omit && \omit && \omit && \omit && \omit &\cr
\tablerule
\omit & height2pt & \omit && \omit && \omit && \omit && \omit && \omit &\cr
&& 2 && $\eta^{\alpha (\rho} \; \eta^{\sigma) \beta}$ && 9 && $t^{(
\alpha} \; \eta^{\beta) (\rho} \; t^{\sigma)}$ && 16 && $t^{
\alpha} \; t^{\beta} \; x^{\rho} \; x^{\sigma}$ &\cr
\omit & height2pt & \omit && \omit && \omit && \omit && \omit && \omit &\cr
\tablerule
\omit & height2pt & \omit && \omit && \omit && \omit && \omit && \omit &\cr
&& 3 && $\eta^{\alpha\beta} \; t^{\rho} \; t^{\sigma}$ && 10 &&
$t^{(\alpha} \; \eta^{\beta) (\rho} \; x^{\sigma)}$ && 17 && 
$x^{\alpha} \; x^{\beta} \; t^{\rho} \; t^{\sigma}$ &\cr
\omit & height2pt & \omit && \omit && \omit && \omit && \omit && \omit &\cr
\tablerule
\omit & height2pt & \omit && \omit && \omit && \omit && \omit && \omit &\cr
&& 4 && $t^{\alpha} \; t^{\beta} \; \eta^{\rho\sigma}$ && 11 &&
$x^{(\alpha} \; \eta^{\beta) (\rho} \; t^{\sigma)}$ && 18 && 
$t^{(\alpha} \; x^{\beta)} \; t^{(\rho} \; x^{\sigma)}$
&\cr
\omit & height2pt & \omit && \omit && \omit && \omit && \omit && \omit &\cr
\tablerule
\omit & height2pt & \omit && \omit && \omit && \omit && \omit && \omit &\cr
&& 5 && $\eta^{\alpha\beta} \; t^{(\rho} \; x^{\sigma)}$ && 12 
&& $x^{(\alpha} \; \eta^{\beta) (\rho} \; x^{\sigma)}$ && 19 &&
$t^{(\alpha} \; x^{\beta)} \; x^{\rho} \; x^{\sigma}$ 
&\cr
\omit & height2pt & \omit && \omit && \omit && \omit && \omit && \omit &\cr
\tablerule
\omit & height2pt & \omit && \omit && \omit && \omit && \omit && \omit &\cr
&& 6 && $t^{(\alpha} \; x^{\beta)} \; \eta^{\rho\sigma}$ && 13 
&& $t^{\alpha} \; t^{\beta} \; t^{\rho} \; t^{\sigma}$ 
&& 20 && $x^{\alpha} \; x^{\beta} \; t^{(\rho} \; 
x^{\sigma)}$ &\cr
\omit & height2pt & \omit && \omit && \omit && \omit && \omit && \omit &\cr
\tablerule
\omit & height2pt & \omit && \omit && \omit && \omit && \omit && \omit &\cr
&& 7 && $\eta^{\alpha\beta} \; x^{\rho} \; x^{\sigma}$ && 14 &&
$t^{\alpha} \; t^{\beta} \; t^{(\rho} \; x^{\sigma)}$ 
&& 21 && $x^{\alpha} \; x^{\beta} \; x^{\rho} \; 
x^{\sigma}$ &\cr
\omit & height2pt & \omit && \omit && \omit && \omit && \omit && \omit &\cr
\tablerule}}

{\bf Table~3:} {\ninepoint Tensor factors in the self-energy.}

It is worth working out an example to illustrate the method. To save space
during the contractions we define the normal and log propagators as:
$$i\Delta_N \equiv {H^2 \over 8 \pi^2} \; {2 u' u'' \over x^2 + i 
\epsilon} \qquad , \qquad i\Delta_L \equiv {H^2 \over 8 \pi^2} \; \ln\Bigl[H^2 
(x^2 + i \epsilon)\Bigr] \eqno(2.24)$$
Now consider the $i=j=41$ term from the symmetric graviton loop (Fig.~4a):
$$\eqalignno{\Bigl[{^{\alpha_1\beta_1}}\Sigma_{41-41}^{\rho_1\sigma_1}\Bigr](
x';x'') = -i V_{41}^{\alpha_1\beta_1\alpha_2\beta_2\alpha_3\beta_3}(&x';
\partial_1',\partial_2',\partial_3') \enskip i\Bigl[{_{\alpha_2\beta_2}}
\Delta_{\rho_2\sigma_2}\Bigr](x';x'') \cr
\times i\Bigl[{_{\alpha_3\beta_3}} \Delta_{\rho_3\sigma_3}\Bigr]&(x';
x'') \enskip V_{41}^{\rho_1\sigma_1\rho_2\sigma_2\rho_3\sigma_3}(x'';
\partial_1'',\partial_2'',\partial_3'') \qquad \qquad \qquad &(2.25a) \cr}$$
$$\eqalignno{=&-i \kappa^2  \; {\Omega'}^2 \; {\Omega''}^2 
\; \partial_1' \cdot \partial_3' \enskip \partial_1'' \cdot 
\partial_3'' \enskip \eta^{\alpha_1) (\alpha_2} \; \eta^{\beta_2) 
(\alpha_3} \; \eta^{\beta_3) (\beta_1} \enskip \eta^{\rho_1) (\rho_2} 
\; \eta^{\sigma_2) (\rho_3} \; \eta^{\sigma_3) (\sigma_1} \times \cr
&\times \Bigl\{ i\Delta_{2N} \; \Bigl[2 \eta_{\alpha_2 (\rho_2} 
\; \eta_{\sigma_2) \beta_2} - \eta_{\alpha_2 \beta_2} \; 
\eta_{\rho_2 \sigma_2}\Bigr] - i\Delta_{2L} \; \Bigl[2 {\overline 
\eta}_{\alpha_2 (\rho_2} \; {\overline \eta}_{\sigma_2) \beta_2} -2 
{\overline \eta}_{\alpha_2 \beta_2} \; {\overline \eta}_{\rho_2 
\sigma_2}\Bigr] \Bigr\} \cr
&\times \Bigl\{ i\Delta_{3N} \; \Bigl[2 \eta_{\alpha_3 (\rho_3} 
\; \eta_{\sigma_3) \beta_3} - \eta_{\alpha_3 \beta_3} \; 
\eta_{\rho_3 \sigma_3}\Bigr] - i\Delta_{3L} \; \Bigl[2 {\overline 
\eta}_{\alpha_3 (\rho_3} \; {\overline \eta}_{\sigma_3) \beta_3} -2
{\overline \eta}_{\alpha_3 \beta_3} \; {\overline \eta}_{\rho_3 
\sigma_3}\Bigr] \Bigr\} \qquad \qquad &(2.25b)\cr
=&-i \kappa^2  \; {\Omega'}^2 \; {\Omega''}^2 \;
\partial_1' \cdot \partial_3' \enskip \partial_1'' \cdot \partial_3'' \cr
&\times \Biggl\{i\Delta_{2N} \; i\Delta_{3N} \; \Bigl[2 \eta^{
\alpha_1\beta_1} \; \eta^{\rho_1 \sigma_1} + 2 \eta^{\alpha_1 (\rho_1}
\; \eta^{\sigma_1) \beta_1}\Bigr] + \Bigl(i\Delta_{2N} \; 
i\Delta_{3L} + i\Delta_{2L} \; i\Delta_{3N}\Bigr) \times \cr
&\quad \times \Bigl[-3 \eta^{\alpha_1\beta_1} \; \eta^{\rho_1\sigma_1}
+ \eta^{\alpha_1 (\rho_1} \; \eta^{\sigma_1) \beta_1} -3 \eta^{\alpha_1
\beta_1} t^{\rho_1} t^{\sigma_1} - 3 t^{\alpha_1} t^{\beta_1} \; 
\eta^{\rho_1 \sigma_1} + 4 t^{(\alpha_1} \eta^{\beta_1) (\rho_1} t^{\sigma_1)}
\Bigr] \cr
&\qquad + i\Delta_{2L} \; i\Delta_{3L} \; \Bigl[ 5 \eta^{
\alpha_1\beta_1} \; \eta^{\rho_1\sigma_1} -3 \eta^{\alpha_1 (\rho_1} 
\; \eta^{\sigma_1) \beta_1} + 5 \eta^{\alpha_1 \beta_1} t^{\rho_1} 
t^{\sigma_1} \cr
&\qquad \qquad \qquad \qquad \qquad + 5 t^{\alpha_1} t^{\beta_1} \eta^{\rho_1 
\sigma_1} - 6 t^{(\alpha_1} \eta^{\beta_1) (\rho_1} t^{\sigma_1)} + 2 t^{
\alpha_1} t^{\beta_1} t^{\rho_1} t^{\sigma_1} \Bigr] \Biggr\} \qquad \qquad
&(2.25c) \cr}$$
We now label the various tensor factors according to the scheme of Table~3 as
$T_1 \equiv \eta^{\alpha_1 \beta_1} \; \eta^{\rho_1 \sigma_1}$ etc., 
and act the inner derivatives: 
$$\eqalignno{&\Bigl[{^{\alpha_1\beta_1}}\Sigma_{41-41}^{\rho_1\sigma_1}\Bigr](
x';x'') =-i \kappa^2  \; {\Omega'}^2 \; {\Omega''}^2 
\; \partial_1' \cdot \partial_3' \enskip \partial_1'' \cdot 
\partial_3'' \Biggl\{i\Delta_{2N} \; i\Delta_{3N} \; 
\Bigl[2 T_1 + 2 T_2 \Bigr] \cr
&\qquad \qquad \qquad + \Bigl(i\Delta_{2N} \; i\Delta_{3L} + 
i\Delta_{2L} \; i\Delta_{3N}\Bigr) \; \Bigl[-3 T_1 + T_2 - 3 
T_3 - 3 T_4 + 4 T_9 \Bigr] \cr
&\qquad \qquad \qquad \qquad \qquad + i\Delta_{2L} \; i\Delta_{3L} 
\; \Bigl[ 5 T_1 - 3 T_2 + 5 T_3 + 5 T_4 - 6 T_9 + 2 T_{13} \Bigr] 
\Biggr\} &(2.26a) \cr
&={-i \kappa^2 \over 2^6 \pi^4} \Biggl\{ \Bigl(-\frac{32 x\cdot\partial_1' x
\cdot\partial_1''}{x^8} + \frac{8 \partial_1' \cdot \partial_1''}{x^6} + 
\frac{8 x \cdot \partial_1' t \cdot \partial_1''}{u'' x^6} - \frac{8 t \cdot 
\partial_1' x\cdot \partial_1''}{u' x^6} + \frac{4 t \cdot \partial_1' t \cdot
\partial_1''}{u' u'' x^4}\Bigr) \; \Bigl[2 T_1 + 2 T_2 \Bigr] \cr
&\qquad + \Bigl[ \Bigl(-\frac{16 x \cdot \partial_1' x 
\cdot \partial_1''}{u' u'' x^6} + \frac{4 \partial_1' \cdot \partial_1''}{u' 
u'' x^4} + \frac{4 x \cdot \partial_1' t \cdot \partial_1''}{u' {u''}^2 x^4} - 
\frac{4 t \cdot \partial_1' x \cdot \partial_1''}{{u'}^2 u'' x^4} + \frac{2 t
\cdot \partial_1' t \cdot \partial_1''}{{u'}^2 {u''}^2 x^2}\Bigr) \; 
\ln({\scriptstyle H^2 x^2}) \cr
&\qquad \qquad + \Bigl(-\frac{4 \partial_1' \cdot \partial_1''}{u' u'' x^4} + 
\frac{8 x \cdot \partial_1' x \cdot \partial_1''}{u' u'' x^6}\Bigr) \Bigl]
\; \Bigl[-3 T_1 + T_2 - 3 T_3 - 3 T_4 + 4 T_9 \Bigr] \cr
&+ \Bigl(-\frac{2 \partial_1' \cdot \partial_1''}{{u'}^2 {u''}^2 x^2} + 
\frac{4 x \cdot \partial_1' x \cdot \partial_1''}{{u'}^2 {u''}^2 x^4}\Bigr) 
\; \ln({\scriptstyle H^2 x^2}) \; \Bigl[ 5 T_1 - 3 T_2 + 
5 T_3 + 5 T_4 - 6 T_9 + 2 T_{13} \Bigr] \Biggl\} &(2.26b) \cr}$$
This is the point at which the outer vertex and propagators would be attached
in computing the two loop tadpole. However, to extract the one loop self we 
partially integrate the outer derivatives to act them back on $x'$ and $x''$
using (2.23). The result is:
$$\eqalignno{&\Bigl[{^{\alpha_1\beta_1}}\Sigma_{41-41}^{\rho_1\sigma_1}\Bigr](
x';x'') ={-i \kappa^2 \over 2^6 \pi^4} \Biggl\{ \Bigl(\frac{192}{x^8} + \frac{
32}{u' u'' x^6} + \frac{4}{{u'}^2 {u''}^2 x^4}\Bigr) \; \Bigl[2 T_1 + 
2 T_2 \Bigr] \cr
&+ \Bigl(- \frac{32}{u' u'' x^6} - \frac{8 {\Delta u}^2}{{u'}^3 {u''}^3 x^4} - 
\frac{28}{{u'}^2 {u''}^2 x^4} + \frac{4 \ln(H^2 x^2)}{{u'}^2 {u''}^2 x^4} + 
\frac{8 \ln(H^2 x^2)}{{u'}^3 {u''}^3 x^2} \; \Bigl[-3 T_1 + T_2 - 
3 T_3 - 3 T_4 + 4 T_9 \Bigr] \cr
&+ \Bigl(-\frac{8 {\Delta u}^2}{{u'}^3 {u''}^3 x^4} - \frac{16}{{u'}^2 {u''}^2 
x^4} + \frac{8 \ln(H^2 x^2)}{{u'}^3 {u''}^3 x^2} \Bigr) \; \Bigl[5 T_1 
- 3 T_2 + 5 T_3 + 5 T_4 - 6 T_9 + 2 T_{13} \Bigr] \Biggr\} &(2.27) \cr}$$
Note that ${\Delta u} \equiv u'' - u'$ and that we have suppressed the factors
of $i\epsilon$ which go with $x^2$'s. The final results for the entire 
self-energy are given in Tables~4a and 4b.
\vskip .5cm
\vbox{\tabskip=0pt \offinterlineskip
\def\tablerule{\noalign{\hrule}}
\halign to485pt {\strut#& \vrule#\tabskip=1em plus2em& \hfil#& \vrule#& 
\hfil#\hfil& \vrule#\tabskip=0pt\cr
\tablerule
\omit & height2pt & \omit && \omit &\cr
&& \omit\hidewidth {\rm \#}\hidewidth&& \omit\hidewidth {\rm Coefficient}
\hidewidth&\cr
\omit & height2pt & \omit && \omit &\cr
\tablerule
\omit & height2pt & \omit && \omit &\cr
&& 1 && $\frac{1056}{x^8} - \frac{96 {u'}^2}{{u''}^2 x^8} - \frac{288 u'}{
u'' x^8} - \frac{288 u''}{u' x^8} - \frac{96 {u''}^2}{{u'}^2 x^8} - \frac{48}{
{u'}^2 x^6} - \frac{64 u'}{{u''}^3 x^6} - \frac{48}{{u''}^2 x^6}$ &\cr
&& \omit && $+ \frac{132}{u' u'' x^6} - \frac{64 u''}{{u'}^3 x^6} - \frac{40}{
u' {u''}^3 x^4} + \frac{41}{{u'}^2 {u''}^2 x^4} - \frac{40}{{u'}^3 u'' x^4}$ &
\cr\omit & height2pt & \omit && \omit &\cr
\tablerule
\omit & height2pt & \omit && \omit &\cr
&& 2 && $\frac{1808}{x^8} - \frac{544 u'}{u'' x^8} - \frac{544 u''}{u' x^8} - 
\frac{176}{{u'}^2 x^6} + \frac{32 u'}{{u''}^3 x^6} - \frac{176}{{u''}^2 x^6} + 
\frac{272}{u' u'' x^6}$ &\cr
&& \omit && $+ \frac{32 u''}{{u'}^3 x^6} + \frac{24}{u' {u''}^3 x^4}
+ \frac{28}{{u'}^2 {u''}^2 x^4} + \frac{24}{{u'}^3 u'' x^4}$ &\cr
\omit & height2pt & \omit && \omit &\cr
\tablerule
\omit & height2pt & \omit && \omit &\cr
&& 3 && $- \frac{112}{{u'}^2 x^6} + \frac{32 u'}{{u''}^3 x^6} - \frac{16}{{u''
}^2 x^6} - \frac{640}{u' u'' x^6} - \frac{8}{u' {u''}^3 x^4} - \frac{102}{
{u'}^2 {u''}^2 x^4} - \frac{80}{{u'}^3 u'' x^4}$ &\cr
\omit & height2pt & \omit && \omit &\cr
\tablerule
\omit & height2pt & \omit && \omit &\cr
&& 4 && $- \frac{16}{{u'}^2 x^6} - \frac{112}{{u''}^2 x^6} - \frac{640}{u' u'' 
x^6} + \frac{32 u''}{{u'}^3 x^6} - \frac{80}{u' {u''}^3 x^4} - \frac{102}{
{u'}^2 {u''}^2 x^4} - \frac{8}{{u'}^3 u'' x^4}$ &\cr
\omit & height2pt & \omit && \omit &\cr
\tablerule
\omit & height2pt & \omit && \omit &\cr
&& 5 && $ \frac{1344}{u' x^8} - \frac{576 u'}{{u''}^2 x^8} + \frac{928}{u'' x^8}
+ \frac{96 u''}{{u'}^2 x^8} + \frac{64}{{u'}^3 x^6} - \frac{64}{{u''}^3 x^6} + 
\frac{64}{u' {u''}^2 x^6} + \frac{168}{{u}'^2 u'' x^6} - \frac{24}{{u'}^3 
{u''}^2 x^4}$ &\cr
\omit & height2pt & \omit && \omit &\cr
\tablerule
\omit & height2pt & \omit && \omit &\cr
&& 6 && $ -\frac{928}{u' x^8} - \frac{96 u'}{{u''}^2 x^8} - \frac{1344}{u'' 
x^8} + \frac{576 u''}{{u'}^2 x^8} + \frac{64}{{u'}^3 x^6} - \frac{64}{{u''}^3 
x^6} - \frac{168}{u' {u''}^2 x^6} - \frac{64}{{u'}^2 u'' x^6} + \frac{24}{{u'}^2
{u''}^3 x^4}$ &\cr
\omit & height2pt & \omit && \omit &\cr
\tablerule
\omit & height2pt & \omit && \omit &\cr
&& 7 && $ -\frac{1280}{x^{10}} + \frac{1152 u'}{u'' x^{10}} - \frac{704 u''}{u' 
x^{10}} + \frac{8}{{u'}^2 x^8} + \frac{192}{{u''}^2 x^8}- \frac{144}{u' u'' x^8}
+ \frac{32}{u' {u''}^3 x^6} - \frac{24}{{u'}^2 {u''}^2 x^6} + \frac{64}{{u'}^3 
u'' x^6}$ &\cr
\omit & height2pt & \omit && \omit &\cr
\tablerule
\omit & height2pt & \omit && \omit &\cr
&& 8 && $ -\frac{1280}{x^{10}} - \frac{704 u'}{u'' x^{10}} + \frac{1152 u''}{u' 
x^{10}}+ \frac{192}{{u'}^2 x^8} + \frac{8}{{u''}^2 x^8} - \frac{144}{u' u'' x^8}
+ \frac{64}{u' {u''}^3 x^6} - \frac{24}{{u'}^2 {u''}^2 x^6} + \frac{32}{{u'}^3
u'' x^6}$ &\cr
\omit & height2pt & \omit && \omit &\cr
\tablerule
\omit & height2pt & \omit && \omit &\cr
&& 9 && $ \frac{96}{{u'}^2 x^6} + \frac{96}{{u''}^2 x^6} - \frac{464}{u' u'' 
x^6} + \frac{56}{u' {u''}^3 x^4} + \frac{8}{{u'}^2 {u''}^2 x^4} + \frac{56}{
{u'}^3 u'' x^4}$ &\cr
\omit & height2pt & \omit && \omit &\cr
\tablerule
\omit & height2pt & \omit && \omit &\cr
&& 10 && $ \frac{224}{u' x^8} - \frac{1024}{u'' x^8} + \frac{128 u''}{{u'}^2 
x^8} - \frac{64}{{u'}^3 x^6} + \frac{64}{{u''}^3 x^6} - \frac{360}{u' {u''}^2 
x^6} + \frac{120}{{u'}^2 u'' x^6} - \frac{8}{{u'}^2 {u''}^3 x^4}$ &\cr
\omit & height2pt & \omit && \omit &\cr
\tablerule
\omit & height2pt & \omit && \omit &\cr
&& 11 && $ \frac{1024}{u' x^8} - \frac{128 u'}{{u''}^2 x^8} - \frac{224}{u'' 
x^8} - \frac{64}{{u'}^3 x^6} + \frac{64}{{u''}^3 x^6} - \frac{120}{u' {u''}^2 
x^6} + \frac{360}{{u'}^2 u'' x^6} + \frac{8}{{u'}^3 {u''}^2 x^4}$ &\cr
\omit & height2pt & \omit && \omit &\cr
\tablerule
\omit & height2pt & \omit && \omit &\cr
&& 12 && $ -\frac{6528}{x^{10}} + \frac{1408 u'}{u'' x^{10}}+\frac{1408 u''}{u' 
x^{10}} + \frac{160}{{u'}^2 x^8} + \frac{160}{{u''}^2 x^8} - \frac{416}{u' u'' 
x^8} - \frac{64}{u' {u''}^3 x^6} - \frac{24}{{u'}^2 {u''}^2 x^6} - \frac{64}{
{u'}^3 u'' x^6}$ &\cr
\omit & height2pt & \omit && \omit &\cr
\tablerule
\omit & height2pt & \omit && \omit &\cr
&& 13 && $ \frac{40}{u' {u''}^3 x^4} + \frac{180}{{u'}^2 {u''}^2 x^4} + \frac{
40}{{u'}^3 u'' x^4}$ &\cr
\omit & height2pt & \omit && \omit &\cr
\tablerule
\omit & height2pt & \omit && \omit &\cr
&& 14 && $ -\frac{32}{{u'}^3 x^6} - \frac{640}{u' {u''}^2 x^6} - \frac{400}{
{u'}^2 u'' x^6} - \frac{8}{{u'}^2 {u''}^3 x^4} + \frac{8}{{u'}^3 {u''}^2 x^4}$ 
&\cr \omit & height2pt & \omit && \omit &\cr
\tablerule
\omit & height2pt & \omit && \omit &\cr
&& 15 && $ \frac{32}{{u''}^3 x^6} + \frac{400}{u' {u''}^2 x^6} + \frac{640}{
{u'}^2 u'' x^6} - \frac{8}{{u'}^2 {u''}^3 x^4} + \frac{8}{{u'}^3 {u''}^2 x^4}$
&\cr \omit & height2pt & \omit && \omit &\cr
\tablerule
\omit & height2pt & \omit && \omit &\cr
&& 16 && $ \frac{240}{{u'}^2 x^8} + \frac{992}{u' u'' x^8} + \frac{32}{u' 
{u''}^3 x^6} + \frac{56}{{u'}^2 {u''}^2 x^6}$ &\cr
\omit & height2pt & \omit && \omit &\cr
\tablerule
\omit & height2pt & \omit && \omit &\cr
&& 17 && $ \frac{240}{{u''}^2 x^8} + \frac{992}{u' u'' x^8} + \frac{56}{{u'}^2 
{u''}^2 x^6} + \frac{32}{{u'}^3 u'' x^6}$ &\cr
\omit & height2pt & \omit && \omit &\cr
\tablerule
\omit & height2pt & \omit && \omit &\cr
&& 18 && $ -\frac{960}{{u'}^2 x^8} - \frac{960}{{u''}^2 x^8} + \frac{96}{u' u''
x^8} - \frac{64}{u' {u''}^3 x^6} + \frac{96}{{u'}^2 {u''}^2 x^6} - \frac{64}{
{u'}^3 u'' x^6}$ &\cr
\omit & height2pt & \omit && \omit &\cr
\tablerule
\omit & height2pt & \omit && \omit &\cr
&& 19 && $ \frac{3200}{u'' x^{10}} + \frac{576}{u' {u''}^2 x^8} - 
\frac{224}{{u'}^2 u'' x^8}$ &\cr
\omit & height2pt & \omit && \omit &\cr
\tablerule
\omit & height2pt & \omit && \omit &\cr
&& 20 && $ -\frac{3200}{u' x^{10}} + \frac{224}{u' {u''}^2 x^8} - \frac{576}{{u'
}^2 u'' x^8}$ &\cr
\omit & height2pt & \omit && \omit &\cr
\tablerule
\omit & height2pt & \omit && \omit &\cr
&& 21 && $ \frac{5376}{x^{12}}$ &\cr
\omit & height2pt & \omit && \omit &\cr
\tablerule}}

{\bf Table~4a:} {\ninepoint Tensor coefficients which are free of logarithms.
Multiply each term by $-i\kappa^2/(2^6 \pi^4)$ times the
\vskip -8pt \noindent \hglue 2.40 truecm appropriate tensor factor from Table~3
to obtain the contribution to the self-energy.}

\vfill\eject

\vbox{\tabskip=0pt \offinterlineskip
\def\tablerule{\noalign{\hrule}}
\halign to485pt {\strut#& \vrule#\tabskip=1em plus2em& \hfil#& \vrule#& 
\hfil#\hfil& \vrule#\tabskip=0pt\cr
\tablerule
\omit & height2pt & \omit && \omit &\cr
&& \omit\hidewidth {\rm \#}\hidewidth&& \omit\hidewidth {\rm Coefficient}
\hidewidth&\cr
\omit & height2pt & \omit && \omit &\cr
\tablerule
\omit & height2pt & \omit && \omit &\cr
&& 1 && $-\frac{32}{{u'}^2 x^6} + \frac{16 u'}{{u''}^3 x^6} - \frac{32}{{u''}^2
x^6} + \frac{32}{u' u'' x^6} + \frac{16 u''}{{u'}^3 x^6}$ &\cr
&& \omit && $+ \frac{8}{u' {u''}^3 x^4} + \frac{24}{{u'}^2 {u''}^2 x^4} + 
\frac{8}{{u'}^3 u'' x^4} + \frac{12}{{u'}^3 {u''}^3 x^2}$ &\cr
\omit & height2pt & \omit && \omit &\cr
\tablerule
\omit & height2pt & \omit && \omit &\cr
&& 2 && $-\frac{384}{x^8} + \frac{192 u'}{u'' x^8} + \frac{192 u''}{u' x^8} +
\frac{112}{{u'}^2 x^6} - \frac{32 u'}{{u''}^3 x^6} + \frac{112}{{u''}^2 x^6} - 
\frac{128}{u' u'' x^6} - \frac{32 u''}{{u'}^3 x^6}$ &\cr
&& \omit && $-\frac{8}{u' {u''}^3 x^4} - \frac{36}{{u'}^2 {u''}^2 x^4} - 
\frac{8}{{u'}^3 u'' x^4} - \frac{4}{{u'}^3 {u''}^3 x^2}$ &\cr
\omit & height2pt & \omit && \omit &\cr
\tablerule
\omit & height2pt & \omit && \omit &\cr
&& 3 && $\frac{32}{{u'}^2 x^6} + \frac{8}{u' {u''}^3 x^4} - \frac{20}{{u'}^2 
{u''}^2 x^4} + \frac{48}{{u'}^3 u'' x^4} - \frac{4}{{u'}^3 {u''}^3 x^2}$ &\cr
\omit & height2pt & \omit && \omit &\cr
\tablerule
\omit & height2pt & \omit && \omit &\cr
&& 4 && $\frac{32}{{u''}^2 x^6} + \frac{48}{u' {u''}^3 x^4} - \frac{20}{{u'}^2
{u''}^2 x^4} + \frac{8}{{u'}^3 u'' x^4} - \frac{4}{{u'}^3 {u''}^3 x^2}$ &\cr
\omit & height2pt & \omit && \omit &\cr
\tablerule
\omit & height2pt & \omit && \omit &\cr
&& 5 && $-\frac{192}{u' x^8} + \frac{192}{u'' x^8} - \frac{32}{{u'}^3 x^6} + 
\frac{32}{{u''}^3 x^6} - \frac{16}{u' {u''}^2 x^6} - \frac{16}{{u'}^2 u'' x^6}
+ \frac{24}{{u'}^3 {u''}^2 x^4}$ &\cr
\omit & height2pt & \omit && \omit &\cr
\tablerule
\omit & height2pt & \omit && \omit &\cr
&& 6 && $-\frac{192}{u' x^8} + \frac{192}{u'' x^8} - \frac{32}{{u'}^3 x^6} + 
\frac{32}{{u''}^3 x^6} + \frac{16}{u' {u''}^2 x^6} + \frac{16}{{u'}^2 u'' x^6}
- \frac{24}{{u'}^2 {u''}^3 x^4}$ &\cr
\omit & height2pt & \omit && \omit &\cr
\tablerule
\omit & height2pt & \omit && \omit &\cr
&& 7 && $-\frac{16}{u' {u''}^3 x^6} - \frac{16}{{u'}^2 {u''}^2 x^6} - 
\frac{32}{{u'}^3 u'' x^6}$ &\cr
\omit & height2pt & \omit && \omit &\cr
\tablerule
\omit & height2pt & \omit && \omit &\cr
&& 8 && $ -\frac{32}{u' {u''}^3 x^6} - \frac{16}{{u'}^2 {u''}^2 x^6} - 
\frac{16}{{u'}^3 u'' x^6}$ &\cr
\omit & height2pt & \omit && \omit &\cr
\tablerule
\omit & height2pt & \omit && \omit &\cr
&& 9 && $\frac{64}{{u'}^2 x^6} + \frac{64}{{u''}^2 x^6} - \frac{64}{u' u'' x^6}
- \frac{24}{u' {u''}^3 x^4} - \frac{32}{{u'}^2 {u''}^2 x^4} - \frac{24}{{u'}^3 
u'' x^4} - \frac{8}{{u'}^3 {u''}^3 x^2}$ &\cr
\omit & height2pt & \omit && \omit &\cr
\tablerule
\omit & height2pt & \omit && \omit &\cr
&& 10 && $\frac{384}{u' x^8} - \frac{192}{u'' x^8} - \frac{192 u''}{{u'}^2 x^8}
+ \frac{32}{{u'}^3 x^6} - \frac{64}{{u''}^3 x^6} + \frac{64}{u' {u''}^2 x^6} + 
\frac{8}{{u'}^2 {u''}^3 x^4}$ &\cr
\omit & height2pt & \omit && \omit &\cr
\tablerule
\omit & height2pt & \omit && \omit &\cr
&& 11 && $\frac{192}{u' x^8} + \frac{192 u'}{{u''}^2 x^8} - \frac{384}{u'' x^8}
+ \frac{64}{{u'}^3 x^6} - \frac{32}{{u''}^3 x^6} - \frac{64}{{u'}^2 u'' x^6} 
- \frac{8}{{u'}^3 {u''}^2 x^4}$ &\cr
\omit & height2pt & \omit && \omit &\cr
\tablerule
\omit & height2pt & \omit && \omit &\cr
&& 12 && $\frac{3072}{x^{10}} - \frac{1536 u'}{u'' x^{10}} - \frac{1536 u''}{
u' x^{10}} - \frac{288}{{u'}^2 x^8} - \frac{288}{{u''}^2 x^8} + \frac{384}{u' 
u'' x^8} + \frac{32}{u' {u''}^3 x^6} + \frac{80}{{u'}^2 {u''}^2 x^6} + 
\frac{32}{{u'}^3 u'' x^6}$ &\cr 
\omit & height2pt & \omit && \omit &\cr
\tablerule
\omit & height2pt & \omit && \omit &\cr
&& 13 && $-\frac{16}{u' {u''}^3 x^4} + \frac{80}{{u'}^2 {u''}^2 x^4} - 
\frac{16}{{u'}^3 u'' x^4}$ &\cr
\omit & height2pt & \omit && \omit &\cr
\tablerule
\omit & height2pt & \omit && \omit &\cr
&& 14 && $\frac{128}{u' {u''}^2 x^6} - \frac{192}{{u'}^2 u'' x^6} + \frac{8}{
{u'}^2 {u''}^3 x^4} - \frac{8}{{u'}^3 {u''}^2 x^4}$ &\cr
\omit & height2pt & \omit && \omit &\cr
\tablerule
\omit & height2pt & \omit && \omit &\cr
&& 15 && $\frac{192}{u' {u''}^2 x^6} - \frac{128}{{u'}^2 u'' x^6} + \frac{8}{
{u'}^2 {u''}^3 x^4} - \frac{8}{{u'}^3 {u''}^2 x^4}$ &\cr
\omit & height2pt & \omit && \omit &\cr
\tablerule
\omit & height2pt & \omit && \omit &\cr
&& 16 && $\frac{192}{{u'}^2 x^8} - \frac{384}{u' u'' x^8} - \frac{32}{u' 
{u''}^3 x^6} - \frac{16}{{u'}^2 {u''}^2 x^6}$ &\cr
\omit & height2pt & \omit && \omit &\cr
\tablerule
\omit & height2pt & \omit && \omit &\cr
&& 17 && $\frac{192}{{u''}^2 x^8} - \frac{384}{u' u'' x^8} - \frac{16}{{u'}^2 
{u''}^2 x^6} - \frac{32}{{u'}^3 u'' x^6}$ &\cr
\omit & height2pt & \omit && \omit &\cr
\tablerule
\omit & height2pt & \omit && \omit &\cr
&& 18 && $\frac{192}{{u'}^2 x^8} + \frac{192}{{u''}^2 x^8} - \frac{1152}{u' u''
x^8} + \frac{32}{u' {u''}^3 x^6} + \frac{32}{{u'}^3 u'' x^6}$ &\cr
\omit & height2pt & \omit && \omit &\cr
\tablerule
\omit & height2pt & \omit && \omit &\cr
&& 19 && $\frac{1536}{u' x^{10}} - \frac{1536}{u'' x^{10}} - \frac{288}{u' 
{u''}^2 x^8} + \frac{96}{{u'}^2 u'' x^8}$ &\cr
\omit & height2pt & \omit && \omit &\cr
\tablerule
\omit & height2pt & \omit && \omit &\cr
&& 20 && $\frac{1536}{u' x^{10}} - \frac{1536}{u'' x^{10}} - \frac{96}{u' 
{u''}^2 x^8} + \frac{288}{{u'}^2 u'' x^8}$ &\cr
\omit & height2pt & \omit && \omit &\cr
\tablerule
\omit & height2pt & \omit && \omit &\cr
&& 21 && $0$ &\cr
\omit & height2pt & \omit && \omit &\cr
\tablerule}}

{\bf Table~4b:} {\ninepoint Tensor coefficients which contain logarithms.
Multiply each term by $-i\kappa^2 \ln(H^2 x^2)/(2^6 \pi^4)$ 
\vskip -8pt \noindent \hglue 2.40 truecm times the appropriate tensor factor 
from Table~3 to obtain the contribution to the self-energy.}

\vfill\eject
\centerline{III. THE FLAT SPACE LIMIT}

Consideration of the classical background (2.11) reveals that flat space can be
recovered by setting the conformal time to:
$$u = \frac1{H} - t \eqno(3.1)$$
and then taking the Hubble constant $H$ to zero.$^{11}$ Note that in this limit
the scale factor $\Omega$ becomes unity, as does the ratio of products of equal
numbers of conformal times. Note finally that the difference of two conformal
times is just minus the same difference of flat space times:
$$x^0 \equiv u' - u'' = \Bigl(\frac1{H} - t'\Bigr) - \Bigl(\frac1{H} - t''
\Bigr) = t'' - t' \eqno(3.2)$$
This means that our quantity $x^2 \equiv (x'-x'')^{\mu} (x'-x'')^{\nu} \eta_{
\mu\nu}$ goes to the usual Lorentz invariant interval, which we shall continue
to call $x^2$.

The preceding facts make it very simple to take the self-energy's flat space 
limit. Consider, for example, the no-log coefficient of tensor factor \#7, 
which we can read from Table~4a. A simple calculation gives the following 
limit:
$$\eqalignno{-\frac{1280}{x^{10}} + \frac{1152 u'}{u'' x^{10}} - \frac{704 
u''}{u' x^{10}} + \frac{8}{{u'}^2 x^8} + \frac{192}{{u''}^2 x^8}&- \frac{144}{u'
u'' x^8} + \frac{32}{u' {u''}^3 x^6} - \frac{24}{{u'}^2 {u''}^2 x^6} + \frac{
64}{{u'}^3 u'' x^6} \cr
&\longrightarrow -\frac{1280}{x^{10}} + \frac{1152}{x^{10}} - \frac{704}{x^{10}}
= -\frac{832}{x^{10}} &(3.3) \cr}$$
The logarithm terms of Table~4b cancel completely, and the only non-zero
contributions from the non-log terms of Table~4a come from the coefficients of
Tensors \#1, 2, 7, 8, 12, and 21. The answer is:
$$\eqalignno{\Bigl[{^{\alpha\beta}}\Sigma_{\rm flat}^{\rho\sigma}\Bigr](
x';x'') ={-i \kappa^2 \over 2^6 \pi^4} \Biggl\{ \frac{288}{x^8} \;
\eta^{\alpha\beta} \; \eta^{\rho\sigma} + \frac{720}{x^8} &\;
\eta^{\alpha (\rho} \; \eta^{\sigma) \beta} - \frac{832}{x^{10}} 
\; \Bigl[\eta^{\alpha\beta} x^{\rho} x^{\sigma} + x^{\alpha} x^{\beta} 
\eta^{\rho\sigma}\Bigr] \cr
&- \frac{3712}{x^{10}} \; x^{(\alpha} \eta^{\beta) (\rho} x^{\sigma)} +
\frac{5376}{x^{12}} \; x^{\alpha} x^{\beta} x^{\rho} x^{\sigma} 
\Biggr\} \qquad &(3.4) \cr}$$
It is useful to recast this expression in the form of a derivative operator
acting on $1/x^4$. The key identities are:
$$\frac1{x^8} = \frac1{192} \; \partial^4 \; \Bigl(\frac1{x^4}
\Bigr) \eqno(3.5a)$$
$$\frac{x^{\mu} x^{\nu}}{x^{10}} = \Bigl[\frac1{384} \; \partial^{\mu}
\partial^{\nu} \partial^2 + \frac1{1536} \; \eta^{\mu\nu} \;
\partial^4\Bigr] \; \Bigl(\frac1{x^4}\Bigr) \eqno(3.5b)$$
$$\frac{x^{\mu} x^{\nu} x^{\rho} x^{\sigma}}{x^{12}} = \Bigl[\frac1{1920} 
\; \partial^{\mu} \partial^{\nu} \partial^{\rho} \partial^{\sigma} +
\frac1{640} \; \eta^{(\mu \nu} \; \partial^{\rho} \partial^{
\sigma)} \; \partial^2 + \frac1{5120} \; \eta^{(\mu\nu} 
\; \eta^{\rho\sigma)} \; \partial^4\Bigr] \; \Bigl(
\frac1{x^4}\Bigr) \eqno(3.5c)$$
Our final result for the flat space limit of the one loop self-energy is
therefore:
$$\eqalignno{\Bigl[{^{\alpha\beta}}\Sigma_{\rm flat}^{\rho\sigma}\Bigr](
x';x'') ={-i \kappa^2 \over 60 \; (2 \pi)^4} \Biggl\{\frac{23}2 
\; \eta^{\alpha\beta} \; \eta^{\rho\sigma} \; 
\partial^4 &+ \frac{61}2 \; \eta^{\alpha (\rho} \; \eta^{\sigma)
\beta} \; \partial^4 - \frac{23}2 \; \Bigl[\eta^{\alpha \beta}
\; \partial^{\rho} \partial^{\sigma} + \eta^{\rho\sigma} \;
\partial^{\alpha} \partial^{\beta}\Bigr] \; \partial^2 \cr
&- {\scriptstyle 61} \; \partial^{(\alpha} \eta^{\beta) (\rho} 
\partial^{\sigma)} \; \partial^2 + {\scriptstyle 42} \;
\partial^{\alpha} \partial^{\beta} \partial^{\rho} \partial^{\sigma}\Biggr\} 
\; \Bigl(\frac1{x^4}\Bigr) &(3.6) \cr}$$

It remains to compare (3.6) with the flat space result obtained for the 
$\Lambda = 0$ theory by Capper.$^7$ He defined the graviton field 
as:\footnote{*}{\tenpoint Capper also used a timelike metric, he associates the
inverse factors of $2\pi$ in Fourier transforms with what he calls the 
propagators, and what he calls the self-energy is $-i$ times what we call the 
self-energy. We have translated all these conventions into our notation to 
avoid confusion.}
$$g_{\mu\nu} = \eta_{\mu\nu} + \sqrt{2} \kappa \; \phi_{\mu \nu}
\eqno(3.7)$$
and he used a general gauge fixing term of the form:
$${\cal L}_B = - \Bigl(\alpha \phi^{\mu \rho}_{~~,\rho} + \beta \phi^{,\mu}
\Bigr) \; \eta_{\mu \nu} \; \Bigl(\alpha \phi^{\nu \sigma}_{
~~,\sigma} + \beta \phi^{,\nu} \Bigr) \eqno(3.8)$$
Comparison with our gauge fixing term (2.14) implies that our flat space limit
should agree with his result for $\alpha = 1$ and $\beta = -\frac12$. These are
certainly the values for which his graviton and ghost propagators agree with
the flat space limits of (2.19a) and (2.19b) respectively.

Of course Capper worked in momentum space, using dimensional regularization in
$D$ spacetime dimensions to define the divergent loop integral. To compare with
(3.6) we first Fourier transform to position space and then evaluate the result
--- which is well defined for $x^2 \neq 0$ --- at $D=4$:
$$\Bigl[{^{\alpha\beta}} \Sigma^{\rho\sigma}_{\rm Capper}\Bigr](x';x'') = 
\lim_{D \rightarrow 4} \; {i \over (2\pi)^D} \; \int {d^Dp 
\over (2\pi)^D} \; e^{i p \cdot x} \; T^{\alpha\beta 
\rho\sigma}(p^2) \eqno(3.9)$$
Capper's result for $T^{\alpha\beta\rho\sigma}(p^2)$ has the following form:
$$\eqalignno{T^{\alpha\beta\rho\sigma}(p^2) = \kappa^2 \Biggl\{T_1(D) &
\; p^{\alpha} p^{\beta} p^{\rho} p^{\sigma} + T_2(D) \; \eta^{
\alpha\beta} \; \eta^{\rho\sigma} \; p^4 + 2 \; T_3(D) 
\; \eta^{\alpha (\rho} \; \eta^{\sigma) \beta} \; p^4 
&(3.10) \cr
&+ T_4(D) \; \Bigl[\eta^{\alpha\beta} \; p^{\rho} p^{\sigma} +
p^{\alpha} p^{\beta} \; \eta^{\rho\sigma}\Bigr] \; p^2 + 4 
\; T_5(D) \; p^{(\alpha} \eta^{\beta) (\rho} p^{\sigma)} 
\; p^2 \Biggr\} \; I(p^2) \cr}$$
Although Capper's published paper$^7$ quotes only the pole terms, he was kind 
enough to communicate the general result to us some time ago.$^{15}$ The
coefficient functions are:
$$T_1(D) = \frac9{16} D^4 - \frac{21}{16} D^3 - \frac98 D^2 \eqno(3.11a)$$
$$T_2(D) = \frac1{D-2} \Bigl(\frac9{16} D^5 - \frac{39}{16} D^4 - \frac{25}8 
D^3 + \frac{123}8 D^2 + \frac{33}4 D - 8\Bigr) \eqno(3.11b)$$
$$T_3(D) = -T_5(D) = \frac1{D-2} \Bigl(\frac14 D^4 + \frac{17}{16} D^3 - 
\frac{97}{16} D^2 - \frac{17}8 D + 4\Bigr) \eqno(3.11c)$$
$$T_4(D) = \frac1{D-2} \Bigl(-\frac9{16} D^5 + \frac{43}{16} D^4 + \frac{15}8 
D^3 - \frac{119}8 D^2 - \frac{25}4 D + 8\Bigr) \eqno(3.11d)$$
and Capper defines his ``basic integral'' as:
$$I(p^2) \equiv {1 \over 4 (D^2-1)} \int \; {d^Dk \over [k^2 - i 
\epsilon] \; [(p-k)^2 - i \epsilon]} \eqno(3.12)$$
The familiar relation:
$${1 \over k^2 - i \epsilon} = {i \Gamma(\frac{D-2}2) \over 4 \pi^{\frac{D}2}}
\; \int d^Dy \; {e^{-i k \cdot y} \over [y^2 + i \epsilon]^{
\frac{D}2-1}} \eqno(3.13)$$
allows us to re-express Capper's basic integral in the form:
$$I(p^2) = - {\Bigl[\Gamma(\frac{D-2}2)\Bigr]^2 \over 2^{6-D} (D^2-1)} 
\; \int d^Dy \; {e^{-i p \cdot y} \over [y^2 + i 
\epsilon]^{D-2}} \eqno(3.14)$$
Substitution into (3.9) reveals complete agreement with our answer (3.6):
$$\eqalignno{\Bigl[&{^{\alpha\beta}} \Sigma^{\rho\sigma}_{\rm Capper}\Bigr](x';
x'')\cr
&= \lim_{D \rightarrow 4} \; {-i \kappa^2 \Bigl[\Gamma(\frac{D-2}2)
\Bigr]^2 \over 2^6 \pi^D (D^2-1)} \; \Biggl\{T_1(D) \; 
\partial^{\alpha} \partial^{\beta} \partial^{\rho} \partial^{\sigma} + T_2(D) 
\; \eta^{\alpha\beta} \; \eta^{\rho\sigma} \; 
\partial^4 + 2 \; T_3(D) \eta^{\alpha (\rho} \; \eta^{\sigma) 
\beta} \; \partial^4 \cr
&\qquad + T_4(D) \; \Bigl[\eta^{\alpha\beta} \; \partial^{\rho}
\partial^{\sigma} + \partial^{\alpha} \partial^{\beta} \; \eta^{\rho 
\sigma}\Bigr] \; \partial^2 + 4 \; T_5(D) \; \partial^{
(\alpha} \eta^{\beta) (\rho} \partial^{\sigma)} \; \partial^2\Biggr\} 
\; \Bigl(\frac1{x^2 + i \epsilon} \Bigr)^{D-2} &(3.15a) \cr
&= {-i \kappa^2 \over 60 (2 \pi)^4} \; \Biggl\{{\scriptstyle 42} 
\; \partial^{\alpha} \partial^{\beta} \partial^{\rho} \partial^{\sigma}
+ \frac{23}2 \; \eta^{\alpha\beta} \; \eta^{\rho\sigma} 
\; \partial^4 + \frac{61}2 \; \eta^{\alpha (\rho} \; 
\eta^{\sigma) \beta} \; \partial^4 \cr
&\qquad \qquad \qquad \qquad -\frac{23}2 \; 
\Bigl[\eta^{\alpha\beta} \; \partial^{\rho} \partial^{\sigma} + 
\partial^{\alpha} \partial^{\beta} \; \eta^{\rho \sigma}\Bigr] 
\; \partial^2 - {\scriptstyle 61} \; \partial^{(\alpha} 
\eta^{\beta) (\rho} \partial^{\sigma)} \; \partial^2 \Biggr\} 
\; \Bigl(\frac1{x^2 + i \epsilon} \Bigr)^2 \qquad &(3.15b) \cr}$$

\vfill\eject

\centerline{IV. THE WARD IDENTITY}

The result of the previous section is reassuring because our reduction 
procedure is the same for all terms. By checking the flat space limit we have 
therefore partially checked even terms which vanish in this limit. However, it 
is conceivable that an error might also vanish in the flat space limit, and it
is important to note that the dominant infrared terms which are of greatest 
interest to us also go to zero in this limit. So an independent check which is
intrinsic to the curved background would be highly desirable. Such a check is
provided by the Ward identity. Since this is a consequence of the theory's 
gauge invariance --- as reflected in the BRS symmetry of the gauge fixed action
--- it can be checked on the curved background without taking the flat space 
limit.

The gauge fixed action is invariant under the following BRS transformation:
$$\delta_{\scriptscriptstyle BRS} \psi_{\mu\nu} = \Bigl[ 2 \; 
{\widetilde g}_{\rho (\mu} \; \partial_{\nu)} + {\widetilde g}_{\mu \nu
, \rho} + \frac2{u} \; {\widetilde g}_{\mu\nu} \; t_{\rho}
\Bigr] \; \omega^{\rho} \; {\delta\zeta} \eqno(41a)$$
$$\delta_{\scriptscriptstyle BRS} {\overline \omega}^{\mu} = - \Omega^{-1} 
F^{\mu} {\delta\zeta} \eqno(4.1b)$$
$$\delta_{\scriptscriptstyle BRS} \omega_{\mu} = \kappa \; \omega_{\mu,
\nu} \; \omega^{\nu} \; {\delta\zeta} \eqno(4.1c)$$
where $\delta\zeta$ is an anti-commuting $\comp$-number constant. An important
consequence is that the BRS transformation of the gauge fixing function 
$F_{\mu}$ is proportional to the anti-ghost equation of motion:
$$\delta_{\scriptscriptstyle BRS} F_{\mu}(x) = \Omega^{-1} {\partial S_{G.F.} 
\over \delta {\overline \omega}^{\mu}(x)} \; {\delta\zeta} \eqno(4.2)$$
The various Slavnov-Taylor identities follow from the BRS invariance of the
functional formalism. One makes a change of variables that is a BRS 
transformation and then collects the variation terms. Since a functional 
integral is independent of the dummy variable of integration, the sum of the 
variations must vanish. Since the action and the measure factor are BRS
invariant, the variations derive entirely from the operator whose in-out matrix
element is being computed.

We do not want to compute another Green's function so we seek an operator whose
BRS variation involves only the pseudo-graviton 2-point function. The desired 
object turns out to be the product of the conformally rescaled anti-ghost field
and the gauge fixing function. Its BRS variation is:
$$\delta_{\scriptscriptstyle BRS} \Bigl[\Omega(x') \; {\overline 
\omega}_{\mu}(x') \; F_{\nu}(x'')\Bigr] = -\Biggl[F_{\mu}(x') \;
F_{\nu}(x'') + \Omega(x') \; {\overline \omega}_{\mu}(x') \;
\Omega^{-1}(x'') \; {\partial S_{G.F.} \over \delta {\overline \omega}^{
\nu}(x'')}\Biggr] \; {\delta\zeta} \eqno(4.3)$$
The functional integral of the second term is a delta function:
$$\eqalignno{\Fint [d\psi] [d{\overline \omega}]& [d\omega] \; 
{\overline \omega}_{\mu}(x') \; {\delta S_{G.F.} \over \delta 
{\overline \omega}^{\nu}(x'')} \; \exp\Bigl[i S_{G.F.}[\psi,{\overline 
\omega},\omega]\Bigr] \cr
&= -i \Fint [d\psi] [d{\overline \omega}] [d\omega] \; {\overline 
\omega}_{\mu}(x') \; {\delta \over \delta {\overline \omega}^{\nu}(
x'')} \; \exp\Bigl[i S_{G.F.}[\psi,{\overline \omega},\omega]\Bigr] 
&(4.4a) \cr
&= -i \eta_{\mu\nu} \delta^4(x'-x'') &(4.4b) \cr}$$
If we write the gauge fixing function as an operator acting on the 
pseudo-graviton field:
$$\eqalignno{F_{\mu}(x) &= \Omega(x) \Bigl[\delta_{\mu}^{~\rho} \partial^{
\sigma} - \frac12 \partial_{\mu} \eta^{\rho\sigma} + \frac2{u} \delta_{\mu}^{
~\rho} t^{\sigma} \Bigr] \; \psi_{\rho\sigma}(x) &(4.5a) \cr
& \equiv {\cal F}_{\mu}^{~\rho\sigma}(x) \; \psi_{\rho\sigma}(x) &(4.5b)
\cr}$$
then the Ward identity can be expressed as follows:
$${\cal F}_{\mu}^{~\alpha\beta}(x') \; {\cal F}_{\nu}^{~\rho\sigma}(x'')
\; \Bigl\langle {\rm out} \Bigl\vert T\Bigl[ \psi_{\alpha\beta}(x') 
\; \psi_{\rho\sigma}(x'')\Bigr] \Bigr\vert {\rm in} \Bigr\rangle = i
\eta_{\mu\nu} \; \delta^4(x'-x'') \eqno(4.6)$$

One of the first things we did after obtaining the pseudo-graviton and ghost
propagators was to verify a somewhat more general version of (4.6) at tree
order.$^{11}$ Work at higher orders is embarrassed by the severe infrared 
divergences of the in-out formalism, which typically prevent matrix elements 
from existing if they involve even one integration over an interaction 
vertex.$^8$ Of course the one loop correction to the in-out 2-point function is
just the double integral of the one loop self-energy against two external 
propagators. If we can amputate the external propagators, the one loop
self-energy which remains will suffer no infrared divergences because it 
involves no integrated interaction vertices. However, we cannot amputate the 
external propagators in (4.6) so simply on account of the gauge fixing 
operators ${\cal F}_{\mu}^{~\rho\sigma}$. These do not commute with the 
pseudo-graviton kinetic operator --- they don't even possess the requisite 
number of free indices to do so. What we must do instead is to reflect the 
gauge fixing operators through the external propagators and then amputate the 
different {\it scalar} propagators which reside on distinct tensor factors.

Two scalar propagators are of interest. The first is that of a massless, 
minimally coupled scalar:
$$i\Delta_A(x';x'') = {H^2 \over 8 \pi^2} \; \Biggl\{ {2 u' u'' \over
x^2 + i \epsilon} - \ln\Bigl[H^2 (x^2 + i \epsilon)\Bigr]\Biggr\} \eqno(4.7a)$$
It corresponds to the $A$ modes of (2.17a), which  can harbor physical graviton
polarizations. The other scalar propagator is that of a massless, conformally
coupled scalar:
$$i\Delta_B(x';x'') = {H^2 \over 8 \pi^2} \; {2 u' u'' \over x^2 + i 
\epsilon} \eqno(4.7b)$$
It corresponds to the $B$ and $C$ modes of (2.17b), which represent constrained
and pure gauge degrees or freedom.\footnote{*}{\tenpoint $A$ modes can also be 
unphysical. In $3+1$ dimensions there are six $A$ modes, of which only two are 
physical, 3 $B$ modes, and a single $C$ mode. The distinction between $B$ and 
$C$ propagators becomes apparent in higher dimensions.$^{9,11}$} The 
corresponding scalar kinetic operators are, respectively:
$$D_A = \Omega^2 \Bigl(\partial^2 + \frac2{u} \partial_0\Bigr) \eqno(4.8a)$$
$$D_B = \Omega^2 \partial^2 \eqno(4.8b)$$
We can express the pseudo-graviton propagator as simple tensor factors times
the two scalar propagators:$^9$
$$i\Bigl[{_{\alpha\beta}} \Delta_{\rho\sigma}\Bigr](x';x'') = i\Delta_A(x';x'')
\; \Bigl[{_{\alpha\beta}} T^A_{\rho\sigma}\Bigr] + i\Delta_B(x';x'')
\; \Biggl\{ \Bigl[{_{\alpha\beta}} T^B_{\rho\sigma}\Bigr] + \Bigl[{
_{\alpha\beta}} T^C_{\rho\sigma}\Bigr]\Biggr\} \eqno(4.9a)$$
$$\Bigl[{_{\alpha\beta}} T^A_{\rho\sigma}\Bigr] = 2 \; {\overline 
\eta}_{\alpha (\rho} \; {\overline \eta}_{\sigma) \beta} - 2 \;
{\overline \eta}_{\alpha\beta} \; {\overline \eta}_{\rho\sigma} 
\eqno(4.9b)$$
$$\Bigl[{_{\alpha\beta}} T^B_{\rho\sigma}\Bigr] = - 4 \; t_{(\alpha}
{\overline \eta}_{\beta) (\rho} t_{\sigma)} \eqno(4.9c)$$
$$\Bigl[{_{\alpha\beta}} T^C_{\rho\sigma}\Bigr] = \Bigl(t_{\alpha} t_{\beta} +
{\overline \eta}_{\alpha\beta}\Bigr) \Bigl(t_{\rho} t_{\sigma} + {\overline
\eta}_{\rho\sigma}\Bigr) \eqno(4.9d)$$
where we remind the reader that $t_{\mu} \equiv \eta_{\mu 0}$ and ${\overline
\eta}_{\mu\nu} \equiv \eta_{\mu\nu} + t_{\mu} t_{\nu}$.

The tensor $t_{\mu}$ and our barring convention permit us to express a free
derivative as purely spatial and temporal derivatives:
$$\partial_{\mu} = {\overline \partial}_{\mu} - t_{\mu} \partial_0 \eqno(4.10)$$
Spatial derivatives reflect through the scalar propagators the same way they do
in flat space:
$${\overline \partial}_{\mu}' \; i\Delta_A(x';x'') = - {\overline 
\partial}_{\mu}'' \; i\Delta_A(x';x'') \eqno(4.11a)$$
$${\overline \partial}_{\mu}' \; i\Delta_B(x';x'') = - {\overline 
\partial}_{\mu}'' \; i\Delta_B(x';x'') \eqno(4.11b)$$
The reflection identities for temporal derivatives follow from the mode 
expansions of the various propagators:$^{11}$
$$\partial_0' \; i\Delta_A(x';x'') = - \Bigl(\partial_0'' - \frac2{u''}
\Bigr) \; i\Delta_B(x';x'') \eqno(4.12a)$$
$$\Bigl(\partial_0' - \frac2{u'}\Bigr) \; i\Delta_B(x';x'') = - 
\partial_0'' \; i\Delta_A(x';x'') \eqno(4.12b)$$
$$\Bigl(\partial_0' - \frac1{u'}\Bigr) \; i\Delta_B(x';x'') = - 
\Bigl(\partial_0'' - \frac1{u''} \Bigr) \; i\Delta_B(x';x'') 
\eqno(4.12c)$$
We reflect the gauge fixing operator through an external propagator by first
contracting into the tensor factors and then exploiting the scalar reflection 
identities:
$$\eqalignno{{\Omega'}^{-1} &\; {\cal F}_{\mu}^{~\alpha\beta}(x')
\; i\Bigl[{_{\alpha\beta}} \Delta_{\rho\sigma}\Bigr](x';x'') = 2 \Bigl[
{\overline \eta}_{\mu (\rho} {\overline \partial}_{\sigma)}' - t_{\mu} 
{\overline \eta}_{\rho\sigma} \partial_0'\Bigr] \; i\Delta_A(x';x'') \cr
&+2 \Bigl[-t_{\mu} t_{(\rho} {\overline \partial}_{\sigma)}' -{\overline
\eta}_{\mu (\rho} t_{\sigma)} \Bigl(\partial_0' - \frac2{u'}\Bigr) + 2 t_{\mu} 
\Bigl(t_{\rho} t_{\sigma} + {\overline \eta}_{\rho\sigma}\Bigr) \Bigl(
\partial_0' - \frac1{u'}\Bigr)\Bigr] \; i\Delta_B(x';x'') \qquad \qquad
&(4.13a) \cr
&= 2 \Bigl[-{\overline \eta}_{\mu (\rho} {\overline \partial}_{\sigma)}'' +
{\overline \eta}_{\mu (\rho} t_{\sigma)} \partial_0''\Bigr] \; 
i\Delta_A(x';x'') \cr
&\qquad + 2 t_{\mu} \Bigl[ t_{(\rho} {\overline \partial}_{\sigma)}''
+ {\overline \eta}_{\rho \sigma} \Bigl(\partial_0'' - \frac2{u''}\Bigr) - 
\Bigl(t_{\rho} t_{\sigma} + {\overline \eta}_{\rho \sigma}\Bigr) \Bigl(
\partial_0'' - \frac1{u''}\Bigr)\Bigr] \; i\Delta_B(x';x'') \qquad
&(4.13b) \cr
&=-2 {\overline \eta}_{\mu (\rho} \partial_{\sigma)}'' \; i\Delta_A(x';
x'') + 2 t_{\mu} \Bigl[t_{(\rho} \partial_{\sigma)}'' - \frac1{u''} \eta_{\rho
\sigma}\Bigr] \; i\Delta_B(x';x'') &(4.13c) \cr}$$
We can therefore amputate the external propagator by acting the scalar operator
$D_A$ when the free index $\mu$ is spatial, and by acting $D_B$ when $\mu$ is
temporal.

The preceding analysis allows us to pass from (4.6) to the following identity 
on the one loop self-energy:
$$\Bigl(\eta_{\mu \alpha} \partial_{\beta}' - t_{\mu} \eta_{\alpha\beta}
\frac1{u'}\Bigr) \; \Bigl(\eta_{\nu \rho} \partial_{\sigma}'' - 
t_{\nu} \eta_{\rho\sigma} \frac1{u''}\Bigr) \; \Bigl[{^{\alpha\beta}} 
\Sigma^{\rho\sigma}\Bigr](x';x'') = 0 \eqno(4.14)$$
Though our derivation was carried out in $3+1$ dimensions it is worth remarking
that the result is valid for any dimension. To check it in $3+1$ dimensions we
first contract and commute the two tensor--differential operators through the 
21 tensors of Table~3. The result is a linear combination of five 2--index 
tensors:
$$\eta_{\mu\nu} \qquad t_{\mu} t_{\nu} \qquad t_{\mu} x_{\nu} \qquad x_{\mu}
t_{\nu} \qquad x_{\mu} x_{\nu} \eqno(4.15)$$
times various scalar differential operators which act on the appropriate scalar
functions of Tables~4a and 4b. Since the coefficient of each tensor must vanish
separately we obtain five identities, although those for $t_{\mu} x_{\nu}$ and
$x_{\mu} t_{\nu}$ follow from each other by the diagram's invariance under 
interchange of the external legs. The various differential operators have been
tabulated at the end of this section, according to the number of the 
coefficient function upon which they act. Needless to say, the identity is 
obeyed.

It is worth working out an example of the process through which the two tensor
differential operators in (4.14) are contracted and commuted through the various
tensor factors which make up the self-energy. We have chosen for this purpose 
tensor \#21 on Table~3. Since $x^{\mu} = (x'-x'')^{\mu}$ we can easily commute 
derivatives through the tensor factor:
$$\eqalignno{\Bigl(\eta_{\mu \alpha}\partial_{\beta}' &- t_{\mu} \eta_{\alpha
\beta} \frac1{u'}\Bigr) \; \Bigl(\eta_{\nu \rho} \partial_{\sigma}'' 
- t_{\nu} \eta_{\rho\sigma} \frac1{u''}\Bigr) \; x^{\alpha} x^{\beta}
x^{\rho} x^{\sigma} \cr
&= \partial_{\rho}' \partial_{\sigma}'' \; x_{\mu} x_{\nu} x^{\rho} 
x^{\sigma} -\frac1{u'} \partial_{\rho}'' \; t_{\mu} x_{\nu} x^{\rho} 
x^2 - \frac1{u''} \partial_{\rho}' \; x_{\mu} t_{\nu} x^{\rho} x^2 - 
\frac1{u' u''} \; t_{\mu} t_{\nu} x^4 &(4.16a) \cr
&= x_{\mu} x_{\nu} \; \Bigl[ x^{\rho} x^{\sigma} \partial_{\rho}'
\partial_{\sigma}'' - 7 x^{\rho} \partial_{\rho}' + 7 x^{\sigma} \partial_{
\sigma}'' - 42\Bigr] \cr
&\qquad + {t_{\mu} x_{\nu} \over u'} \Bigl[-x^2 x^{\rho} \partial_{\rho}'' + 7
x^2\Bigr] + {x_{\mu} t_{\nu} \over u''} \Bigl[- x^2 x^{\rho} \partial_{\rho}'
- 7 x^2 \Bigr] + t_{\mu} t_{\nu} {x^4 \over u' u''} &(4.16b) \cr}$$ 
The coefficient of $t_{\mu} t_{\nu}$ at the end of (4.16b) is already 
recognizable as \#21 on Table~5b, and the coefficient of $\eta^{\mu\nu}$ is
clearly $0$, in agreement with \#21 on Table~5a. For the rest we recall that 
the derivatives can be simplified when the functions they act depend upon 
${x'}^{\mu}$ and ${x''}^{\mu}$ only through $x^{\mu}$, $u'$ and $u''$:
$$\partial_{\mu}' \longrightarrow \partial_{\mu} - t_{\mu} {\partial \over
\partial u'} \eqno(4.17a)$$
$$\partial_{\mu}'' \longrightarrow -\partial_{\mu} - t_{\mu} {\partial \over
\partial u''} \eqno(4.17b)$$
Since the coefficient functions of Tables~4a and 4b actually depend upon $x^{
\mu}$ only through $x^2$ a further reduction can eventually be made:
$$\partial_{\mu} \longrightarrow 2 x_{\mu} {\partial \over \partial x^2} 
\eqno(4.18)$$
The coefficients of $t_{\mu} x_{\nu}$ and $x_{\mu} t_{\nu}$ in (4.16b) are
therefore:
$${1 \over u'} \Bigl[-x^2 x^{\rho} \partial_{\rho}'' + 7 x^2\Bigr] = {1 \over
u'} \Bigl[2 x^4 {\partial \over \partial x^2} + ({t\cdot x}) \; x^2 
{\partial \over \partial u''} + 7 x^2\Bigr] \eqno(4.19a)$$
$${1 \over u''} \Bigl[-x^2 x^{\rho} \partial_{\rho}' - 7 x^2\Bigr] = {1 \over
u''} \Bigl[-2 x^4 {\partial \over \partial x^2} + ({t\cdot x}) \; x^2 
{\partial \over \partial u'} - 7 x^2\Bigr] \eqno(4.19b)$$
Upon noting that $t\cdot x = u'' - u'$ we recognize (4.19a) and (4.19b) as
\#21 in Tables~5c and 5d respectively. We can also write:
$$\eqalignno{x^{\rho} x^{\sigma} \partial_{\rho}' \partial_{\sigma}'' &= 
x^{\rho} x^{\sigma} \Bigl[\partial_{\rho} - t_{\rho} {\partial \over \partial
u'}\Bigr] \Bigl[-2 x_{\sigma} {\partial \over \partial x^2} - t_{\sigma} 
{\partial \over \partial u''}\Bigr] &(4.20a) \cr
&= -4 x^4 {\partial^2 \over \partial x^4} - 2 x^2 {\partial \over \partial x^2}
- 2 ({t\cdot x}) \; x^2 \Bigl({\partial \over \partial u''} - {\partial
\over \partial u'}\Bigr) + (t\cdot x)^2 {\partial \over \partial u'} {\partial 
\over \partial u''} &(4.20b) \cr}$$
Combining with the other terms in the coefficient of $x_{\mu} x_{\nu}$ in 
(4.16b) then results in \#21 of Table~5e.

Note that Capper's flat space Ward identity:$^7$
$$\eta_{\mu (\alpha} \partial_{\beta)}' \; \eta_{\nu (\rho} \partial_{
\sigma)}'' \; \Bigl[{^{\alpha\beta}} \Sigma^{\rho\sigma}_{\rm Capper}
\Bigr](x';x'') = 0 \eqno(4.21)$$
involves only $\eta_{\mu\nu}$ and $x_{\mu} x_{\nu}$. Since the associated 
scalar functions are just $x^2$ to the power fixed by dimensional analysis, 
Capper gets only two scalar relations, each of which is independent of $x$:
$$T_1(D) + T_2(D) + T_3(D) + 2 \; T_4(D) + 3 \; T_5(D) = 0 
\eqno(4.22a)$$
$$T_3(D) + T_5(D) = 0 \eqno(4.22b)$$
(See (3.10) and (3.11) for the $T_i(D)$.) The presence of $t^{\mu}$ gives us
two more distinct tensors, and the four scalar relations which result can 
involve a bewildering number of distinct products of $x^2$, $u'$, $u''$, and 
$\ln[H^2 x^2]$.
\vskip .5cm
\def\dxx{\frac{\partial}{\partial x^2}}
\def\dx2{\frac{\partial^2}{\partial x^4}}
\def\dup{\frac{\partial}{\partial u'}}
\def\dupp{\frac{\partial}{\partial u''}}
\def\x{\scriptstyle x}

\def\delu{\scriptstyle (u'' - u')}
\def\deldu{\scriptstyle (\dupp-\dup)}

\vbox{\tabskip=0pt \offinterlineskip
\def\tablerule{\noalign{\hrule}}
\halign to470pt {\strut#& \vrule#\tabskip=1em plus2em& \hfil#& \vrule#& 
\hfil#\hfil& \vrule#\tabskip=0pt\cr
\tablerule
\omit & height2pt & \omit && \omit &\cr
&& \omit\hidewidth {\rm \#}\hidewidth&& \omit\hidewidth {\rm Operator}
\hidewidth&\cr
\omit & height2pt & \omit && \omit &\cr
\tablerule
\omit & height2pt & \omit && \omit &\cr
&& 1 && $-{\scriptstyle 2} \frac{\partial}{\partial x^2}$ &\cr
\omit & height2pt & \omit && \omit &\cr
\tablerule
\omit & height2pt & \omit && \omit &\cr
&& 2 && $-\frac12 \dup \dupp -{\scriptstyle 5} \dxx - \delu \deldu \dxx - 
{\scriptstyle 2} \x^2 \dx2$ &\cr
\omit & height2pt & \omit && \omit &\cr
\tablerule
\omit & height2pt & \omit && \omit &\cr
&& 3 && $0$ &\cr
\tablerule
\omit & height2pt & \omit && \omit &\cr
&& 4 && $0$ &\cr
\tablerule
\omit & height2pt & \omit && \omit &\cr
&& 5 && $\frac12 \dupp - \delu \dxx$ &\cr
\omit & height2pt & \omit && \omit &\cr
\tablerule
\omit & height2pt & \omit && \omit &\cr
&& 6 && $-\frac12 \dup - \delu \dxx$ &\cr
\omit & height2pt & \omit && \omit &\cr
\tablerule
\omit & height2pt & \omit && \omit &\cr
&& 7 && $-{\scriptstyle 5} - \delu \dupp - {\scriptstyle 2} \x^2 \dxx$ &\cr
\omit & height2pt & \omit && \omit &\cr
\tablerule
\omit & height2pt & \omit && \omit &\cr
&& 8 && $-{\scriptstyle 5} + \delu \dup - {\scriptstyle 2} \x^2 \dxx$ &\cr
\omit & height2pt & \omit && \omit &\cr
\tablerule
\omit & height2pt & \omit && \omit &\cr
&& 9 && $\frac14 \dup \dupp + \frac12 \dxx + \frac12 \delu \deldu \dxx - 
\delu^2 \dx2$ &\cr
\omit & height2pt & \omit && \omit &\cr
\tablerule
\omit & height2pt & \omit && \omit &\cr
&& 10 && $-\frac54 \dup + \frac14 \dupp - \frac14 \delu \dup \dupp - \frac72
\delu \dxx$ &\cr
&& \omit && $-\frac12 \delu^2 \dupp \dxx - \frac12 \x^2 \dup \dxx -
\delu \x^2 \dx2$ &\cr
\omit & height2pt & \omit && \omit &\cr
\tablerule
\omit & height2pt & \omit && \omit &\cr
&& 11 && $-\frac14 \dup + \frac54 \dupp - \frac14 \delu \dup \dupp
- \frac72 \delu \dxx$ &\cr
&& \omit && $+\frac12 \delu^2 \dup \dxx + \frac12 \x^2 \dupp \dxx -
\delu \x^2 \dx2$ &\cr
\omit & height2pt & \omit && \omit &\cr
\tablerule
\omit & height2pt & \omit && \omit &\cr
&& 12 && $-{\scriptstyle 8} - \frac32 \delu \deldu + \frac14 \delu^2 \dup \dupp 
$ &\cr
&& \omit && $ - \frac{13}2 \x^2 \dxx - \frac12 \delu \deldu \x^2 \dxx - 
\x^4 \dx2$ &\cr
\omit & height2pt & \omit && \omit &\cr
\tablerule
\omit & height2pt & \omit && \omit &\cr
&& 13--21 && $0$ &\cr
\tablerule}}

{\bf Table~5a:} {\ninepoint Operator coefficients of $\eta_{\mu\nu}$ in the 
Ward identity. Act each operator on the corresponding
\vskip -8pt \noindent \hglue 3.10 truecm coefficient function in Tables~4a and 
4b, and then sum the results.}

\vfill\eject

\vbox{\tabskip=0pt \offinterlineskip
\def\tablerule{\noalign{\hrule}}
\halign to470pt {\strut#& \vrule#\tabskip=1em plus2em& \hfil#& \vrule#& 
\hfil#\hfil& \vrule#\tabskip=0pt\cr
\tablerule
\omit & height2pt & \omit && \omit &\cr
&& \omit\hidewidth {\rm \#}\hidewidth&& \omit\hidewidth {\rm Operator}
\hidewidth&\cr
\omit & height2pt & \omit && \omit &\cr
\tablerule
\omit & height2pt & \omit && \omit &\cr
&& 1 && $\dup \dupp + \frac{4}{u'} \dupp + \frac{4}{u''} \dup + \frac{16}{u' 
u''}$ &\cr
\omit & height2pt & \omit && \omit &\cr
\tablerule
\omit & height2pt & \omit && \omit &\cr
&& 2 && $\frac12 \dup \dupp + \frac1{u'} \dupp + \frac1{u''} \dup + \frac{4}{u'
u''}$ &\cr
\omit & height2pt & \omit && \omit &\cr
\tablerule
\omit & height2pt & \omit && \omit &\cr
&& 3 && $-\dup \dupp - {\scriptstyle 10} \dxx - \frac{4}{u'} \dupp + 
{\scriptstyle 2} \delu \dup
\dxx - \frac1{u''} \dup + \frac{4}{u' u''} + {\scriptstyle 8} \frac{u''}{u'}
\dxx$ &\cr
\omit & height2pt & \omit && \omit &\cr
\tablerule
\omit & height2pt & \omit && \omit &\cr
&& 4 && $-\dup \dupp - {\scriptstyle 10} \dxx - \frac1{u'} \dupp - 
{\scriptstyle 2} \delu \dupp
\dxx - \frac{4}{u''} \dup - \frac{4}{u' u''} + {\scriptstyle 8} \frac{u'}{u''}
\dxx$ &\cr
\omit & height2pt & \omit && \omit &\cr
\tablerule
\omit & height2pt & \omit && \omit &\cr
&& 5 && $\frac72 \dup - \frac52 \dupp + \frac{14}{u'} + \frac12 \delu \dup 
\dupp - \frac{5}{u''} - \frac{u'}{u''} \dup + {\scriptstyle 2} \frac{u''}{u'}
\dupp + \x^2 \dup \dxx + {\scriptstyle 4} \frac{x^2}{u'} \dxx$ &\cr
\omit & height2pt & \omit && \omit &\cr
\tablerule
\omit & height2pt & \omit && \omit &\cr
&& 6 && $\frac52 \dup - \frac72 \dupp + \frac{5}{u'} + \frac12 \delu \dup \dupp
- \frac{14}{u''} - {\scriptstyle 2} \frac{u'}{u''} \dup + \frac{u''}{u'} \dupp
- \x^2 \dupp \dxx - {\scriptstyle 4} \frac{x^2}{u''} \dxx$ &\cr
\omit & height2pt & \omit && \omit &\cr
\tablerule
\omit & height2pt & \omit && \omit &\cr
&& 7 && $\frac{x^2}{u''} \dup + {\scriptstyle 4} \frac{x^2}{u' u''}$ &\cr
\omit & height2pt & \omit && \omit &\cr
\tablerule
\omit & height2pt & \omit && \omit &\cr
&& 8 && $\frac{x^2}{u'} \dupp + {\scriptstyle 4} \frac{x^2}{u' u''}$ &\cr
\omit & height2pt & \omit && \omit &\cr
\tablerule
\omit & height2pt & \omit && \omit &\cr
&& 9 && $-\frac34 \dup \dupp - {\scriptstyle 7} \dxx - \frac1{u'} \dupp - \delu
\deldu \dxx - \frac1{u''} \dup$ &\cr
&& \omit && $- \frac1{u' u''} + {\scriptstyle 2} \frac{u'}{u''} \dxx + 
{\scriptstyle 2} \frac{u''}{u'} \dxx - \x^2 \dx2$ &\cr
\omit & height2pt & \omit && \omit &\cr
\tablerule
\omit & height2pt & \omit && \omit &\cr
&& 10 && $\frac74 \dup - \frac54 \dupp + \frac{7}{2 u'} + \frac14 \delu \dup
\dupp - \frac{7}{2 u''}$ &\cr
&& \omit && $- \frac12 \frac{u'}{u''} \dup + \frac12 \frac{u''}{u'} \dupp + 
\frac12 \x^2 \dup \dxx + \frac{x^2}{u'} \dxx - \frac{x^2}{u''} \dxx$ &\cr
\omit & height2pt & \omit && \omit &\cr
\tablerule
\omit & height2pt & \omit && \omit &\cr
&& 11 && $\frac54 \dup - \frac74 \dupp + \frac7{2 u'} + \frac14 \delu \dup 
\dupp - \frac{7}{2 u''}$ &\cr
&& \omit && $- \frac12 \frac{u'}{u''} \dup + \frac12 \frac{u''}{u'} \dupp - 
\frac12 \x^2 \dupp \dxx + \frac{x^2}{u'} \dxx - \frac{x^2}{u''} \dxx$ &\cr
\omit & height2pt & \omit && \omit &\cr
\tablerule
\omit & height2pt & \omit && \omit &\cr
&& 12 && $\frac{x^2}{u' u''}$ &\cr
\omit & height2pt & \omit && \omit &\cr
\tablerule
\omit & height2pt & \omit && \omit &\cr
&& 13 && $\dup \dupp + {\scriptstyle 6} \dxx + \frac1{u'} \dupp + {\scriptstyle
2} \delu \deldu \dxx$ &\cr
&& \omit && $- {\scriptstyle 4} \delu^2 \dx2 + \frac1{u''} \dup + \frac1{u' u''}
- {\scriptstyle 2} \frac{u'}{u''} \dxx - {\scriptstyle 2} \frac{u''}{u'} \dxx$ 
&\cr
\omit & height2pt & \omit && \omit &\cr
\tablerule
\omit & height2pt & \omit && \omit &\cr
&& 14 && $-\frac72 \dup + \frac32 \dupp - \frac{7}{2 u'} - \frac12 \delu \dup
\dupp + {\scriptstyle 12 u'} \dxx - \delu^2 \dupp \dxx + \frac{2}{u''}
+ \frac{u'}{u''} \dup$ &\cr
&& \omit && $ - {\scriptstyle 2} \frac{{u'}^2}{u''} \dxx - {\scriptstyle 10 u''}
\dxx - \frac12 \frac{u''}{u'} \dupp - \x^2 \dup \dxx - \frac{x^2}{u'} \dxx - 
{\scriptstyle 2} \delu \x^2 \dx2$ &\cr
\omit & height2pt & \omit && \omit &\cr
\tablerule
\omit & height2pt & \omit && \omit &\cr
&& 15 && $-\frac32 \dup + \frac72 \dupp - \frac{2}{u'} - \frac12 \delu \dup 
\dupp + {\scriptstyle 10 u'} \dxx + \delu^2 \dup \dxx + \frac7{2 u''} +
\frac12 \frac{u'}{u''} \dup$ &\cr
&& \omit && $+ {\scriptstyle 12 u''} \dxx - \frac{u''}{u'} \dupp + {\scriptstyle
2} \frac{{u''}^2}{u'} \dxx + \x^2 \dupp \dxx - {\scriptstyle 2} \delu \x^2 
\dx2 + \frac{x^2}{u''} \dxx$ &\cr
\omit & height2pt & \omit && \omit &\cr
\tablerule
\omit & height2pt & \omit && \omit &\cr
&& 16 && $-{\scriptstyle 7} - \delu \dupp + {\scriptstyle 2} \frac{u'}{u''} -
{\scriptstyle 4} \x^2 \dxx - \frac{x^2}{u''} \dup - \frac{x^2}{u' u''} + 
{\scriptstyle 2} \frac{u'}{u''} \x^2 \dxx$ &\cr
\omit & height2pt & \omit && \omit &\cr
\tablerule
\omit & height2pt & \omit && \omit &\cr
&& 17 && $-{\scriptstyle 7} + \delu \dup + {\scriptstyle 2} \frac{u''}{u'} -
{\scriptstyle 4} \x^2 \dxx - \frac{x^2}{u'} \dupp - \frac{x^2}{u' u''} + 
{\scriptstyle 2} \frac{u''}{u'} \x^2 \dxx$ &\cr
\omit & height2pt & \omit && \omit &\cr
\tablerule
\omit & height2pt & \omit && \omit &\cr
&& 18 && $-{\scriptstyle 16} - \frac52 {\scriptstyle u'} \dup + {\scriptstyle 
2} {\scriptstyle u'} \dupp + \frac14 \delu^2 \dup \dupp + {\scriptstyle 4} 
\frac{u'}{u''} + \frac12 \frac{{u'}^2}{u''} \dup + {\scriptstyle 2 u''} \dup -
\frac52 {\scriptstyle u''} \dupp + {\scriptstyle 4} \frac{u''}{u'}$ &\cr
&& \omit && $+ \frac12 \frac{{u''}^2}{u'} \dupp - \frac{17}2 \x^2 \dxx - 
\frac12 \delu \deldu \x^2 \dxx + \frac{u'}{u''} \x^2 \dxx +
\frac{u''}{u'} \x^2 \dxx - \x^4 \dx2$ &\cr
\omit & height2pt & \omit && \omit &\cr
\tablerule
\omit & height2pt & \omit && \omit &\cr
&& 19 && $\frac12 \frac{u''-u'}{u''} \x^2 \dup - \frac92 \frac{x^2}{u''} + 
\frac{x^2}{u'} - \frac{x^4}{u''} \dxx$ &\cr
\omit & height2pt & \omit && \omit &\cr
\tablerule
\omit & height2pt & \omit && \omit &\cr
&& 20 && $\frac12 \frac{u''-u'}{u'} \x^2 \dupp + \frac92 \frac{x^2}{u'} - 
\frac{x^2}{u''} + \frac{x^4}{u'} \dxx$ &\cr
\omit & height2pt & \omit && \omit &\cr
\tablerule
\omit & height2pt & \omit && \omit &\cr
&& 21 && $\frac{x^4}{u' u''}$ &\cr
\omit & height2pt & \omit && \omit &\cr
\tablerule}}

{\bf Table~5b:} {\ninepoint Operator coefficients of $t_{\mu} t_{\nu}$ in the 
Ward identity. Act each operator on the corresponding
\vskip -8pt \noindent \hglue 3.10 truecm coefficient function in Tables~4a and 
4b, and then sum the results.}

\vfill\eject

\vskip 1cm
\vbox{\tabskip=0pt \offinterlineskip
\def\tablerule{\noalign{\hrule}}
\halign to470pt {\strut#& \vrule#\tabskip=1em plus2em& \hfil#& \vrule#& 
\hfil#\hfil& \vrule#\tabskip=0pt\cr
\tablerule
\omit & height2pt & \omit && \omit &\cr
&& \omit\hidewidth {\rm \#}\hidewidth&& \omit\hidewidth {\rm Operator}
\hidewidth&\cr
\omit & height2pt & \omit && \omit &\cr
\tablerule
\omit & height2pt & \omit && \omit &\cr
&& 1 && ${\scriptstyle 2} \dup \dxx + \frac{8}{u'} \dxx$ &\cr
\omit & height2pt & \omit && \omit &\cr
\tablerule
\omit & height2pt & \omit && \omit &\cr
&& 2 && $- \dupp \dxx + \frac2{u'} \dxx$ &\cr
\omit & height2pt & \omit && \omit &\cr
\tablerule
\omit & height2pt & \omit && \omit &\cr
&& 3 && $0$ &\cr
\omit & height2pt & \omit && \omit &\cr
\tablerule
\omit & height2pt & \omit && \omit &\cr
&& 4 && $- {\scriptstyle 2} \dup \dxx - \frac2{u'} \dxx - {\scriptstyle 4} 
\delu \dx2$ &\cr
\omit & height2pt & \omit && \omit &\cr
\tablerule
\omit & height2pt & \omit && \omit &\cr
&& 5 && $-\frac12 \dup \dupp - {\scriptstyle 5} \dxx - \frac{2}{u'} \dupp +
\delu \dup \dxx + {\scriptstyle 4} \frac{u''}{u'} \dxx$ &\cr
\omit & height2pt & \omit && \omit &\cr
\tablerule
\omit & height2pt & \omit && \omit &\cr
&& 6 && $-{\scriptstyle 9} \dxx + \delu \dup \dxx + \frac{u''}{u'} \dxx - 
{\scriptstyle 2} \x^2 \dx2$ &\cr
\omit & height2pt & \omit && \omit &\cr
\tablerule
\omit & height2pt & \omit && \omit &\cr
&& 7 && $-{\scriptstyle 5} \deldu + \frac{20}{u'} + \delu \dup \dupp + 
{\scriptstyle 4} \frac{u''}{u'} \dupp + {\scriptstyle 2} \x^2 \dup \dxx + 
{\scriptstyle 8} \frac{x^2}{u'} \dxx$ &\cr
\omit & height2pt & \omit && \omit &\cr
\tablerule
\omit & height2pt & \omit && \omit &\cr
&& 8 && $\frac{2}{u'} + {\scriptstyle 2} \frac{x^2}{u'} \dxx$ &\cr
\omit & height2pt & \omit && \omit &\cr
\tablerule
\omit & height2pt & \omit && \omit &\cr
&& 9 && $\frac12 \dupp \dxx - \delu \dx2$ &\cr
\omit & height2pt & \omit && \omit &\cr
\tablerule
\omit & height2pt & \omit && \omit &\cr
&& 10 && $-\frac12 \dup \dupp - \frac{15}2 \dxx - \frac12 \frac1{u'} \dupp
+\frac12 \delu \dup \dxx$ &\cr
&& \omit && $-\frac32 \delu \dupp \dxx + \frac{u''}{u'} \dxx - {\scriptstyle 2} 
\x^2 \dx2$ &\cr
\omit & height2pt & \omit && \omit &\cr
\tablerule
\omit & height2pt & \omit && \omit &\cr
&& 11 && $-\frac32 \dxx - \frac12 \frac1{u'} \dupp + \frac{u''}{u'} \dxx$ &\cr
\omit & height2pt & \omit && \omit &\cr
\tablerule
\omit & height2pt & \omit && \omit &\cr
&& 12 && $\frac12 \dup - \frac52 \dupp + \frac{5}{u'} + \frac14 \delu \dup 
\dupp + \frac{u''}{u'} \dupp - \frac12 \x^2 \dupp \dxx + {\scriptstyle 2}
\frac{x^2}{u'} \dxx$ &\cr
\omit & height2pt & \omit && \omit &\cr
\tablerule
\omit & height2pt & \omit && \omit &\cr
&& 13 && $0$ &\cr
\omit & height2pt & \omit && \omit &\cr
\tablerule
\omit & height2pt & \omit && \omit &\cr
&& 14 && $\frac12 \dup \dupp + {\scriptstyle 2} \dxx + \frac12 \frac1{u'} \dupp
+ \frac12 \delu \deldu \dxx - {\scriptstyle 2} \delu^2 \dx2 - \frac{u''}{u'} 
\dxx$ &\cr
\omit & height2pt & \omit && \omit &\cr
\tablerule
\omit & height2pt & \omit && \omit &\cr
&& 15 && $0$ &\cr
\omit & height2pt & \omit && \omit &\cr
\tablerule
\omit & height2pt & \omit && \omit &\cr
&& 16 && $-{\scriptstyle 5} \dup + {\scriptstyle 2} \dupp - \frac{5}{u'} - 
\delu \dup \dupp - {\scriptstyle 14} \delu \dxx$ &\cr
&& \omit && $- {\scriptstyle 2} \delu^2 \dupp \dxx - \frac{u''}{u'} \dupp - 
{\scriptstyle 2} \x^2 \dup \dxx - {\scriptstyle 2} \frac{x^2}{u'} \dxx - 
{\scriptstyle 4} \delu \x^2 \dx2$ &\cr
\omit & height2pt & \omit && \omit &\cr
\tablerule
\omit & height2pt & \omit && \omit &\cr
&& 17 && $0$ &\cr
\omit & height2pt & \omit && \omit &\cr
\tablerule
\omit & height2pt & \omit && \omit &\cr
&& 18 && $-\frac12 \dup + {\scriptstyle 2} \dupp - \frac12 \frac1{u'} - \frac14
\delu \dup \dupp + \frac{11}2 {\scriptstyle u'} \dxx + \frac12 \delu^2 \dup 
\dxx$ &\cr
&& \omit && $-\frac{13}2 {\scriptstyle u''} \dxx - \frac12 \frac{u''}{u'} \dupp
+ \frac{{u''}^2}{u'} \dxx + \frac12 \x^2 \dupp \dxx - \delu \x^2 \dx2$ &\cr
\omit & height2pt & \omit && \omit &\cr
\tablerule
\omit & height2pt & \omit && \omit &\cr
&& 19 && $-{\scriptstyle 24} + {\scriptstyle 3} \delu \dup + \frac92 
{\scriptstyle u'} \dupp + \frac12 \delu^2 \dup \dupp - \frac{11}2 {\scriptstyle
u''} \dupp + {\scriptstyle 6} \frac{u''}{u'} + \frac{{u''}^2}{u'} \dupp$ &\cr
&& \omit && $-{\scriptstyle 16} \x^2 \dxx - \delu \deldu \x^2 \dxx + 
{\scriptstyle 2} \frac{u''}{u'} \x^2 \dxx - {\scriptstyle 2} \x^4 \dx2$ &\cr
\omit & height2pt & \omit && \omit &\cr
\tablerule
\omit & height2pt & \omit && \omit &\cr
&& 20 && $-{\scriptstyle 4} + \frac12 \delu \dup + \frac{u''}{u'} - 
{\scriptstyle 2} \x^2 \dxx - \frac12 \frac{x^2}{u'} \dupp + \frac{u''}{u'} \x^2
\dxx$ &\cr
\omit & height2pt & \omit && \omit &\cr
\tablerule
\omit & height2pt & \omit && \omit &\cr
&& 21 && $- \x^2 \dupp + \frac{7}{u'} \x^2 + \frac{u''}{u'} \x^2 \dupp + 
{\scriptstyle 2} \frac{x^4}{u'} \dxx$ &\cr
\omit & height2pt & \omit && \omit &\cr
\tablerule}}

{\bf Table~5c:} {\ninepoint Operator coefficients of $t_{\mu} x_{\nu}$ in the 
Ward identity. Act each operator on the corresponding
\vskip -8pt \noindent \hglue 3.10 truecm coefficient function in Tables~4a and 
4b, and then sum the results.}

\vfill\eject

\vskip 1cm
\vbox{\tabskip=0pt \offinterlineskip
\def\tablerule{\noalign{\hrule}}
\halign to470pt {\strut#& \vrule#\tabskip=1em plus2em& \hfil#& \vrule#& 
\hfil#\hfil& \vrule#\tabskip=0pt\cr
\tablerule
\omit & height2pt & \omit && \omit &\cr
&& \omit\hidewidth {\rm \#}\hidewidth&& \omit\hidewidth {\rm Operator}
\hidewidth&\cr
\omit & height2pt & \omit && \omit &\cr
\tablerule
\omit & height2pt & \omit && \omit &\cr
&& 1 && $-{\scriptstyle 2} \dupp \dxx - \frac{8}{u''} \dxx$ &\cr
\omit & height2pt & \omit && \omit &\cr
\tablerule
\omit & height2pt & \omit && \omit &\cr
&& 2 && $\dup \dxx - \frac2{u''} \dxx$ &\cr
\omit & height2pt & \omit && \omit &\cr
\tablerule
\omit & height2pt & \omit && \omit &\cr
&& 3 && ${\scriptstyle 2} \dupp \dxx - {\scriptstyle 4} \delu \dx2 + 
\frac2{u''} \dxx$ &\cr
\omit & height2pt & \omit && \omit &\cr
\tablerule
\omit & height2pt & \omit && \omit &\cr
&& 4 && $0$ &\cr
\omit & height2pt & \omit && \omit &\cr
\tablerule
\omit & height2pt & \omit && \omit &\cr
&& 5 && $-{\scriptstyle 9} \dxx - \delu \dupp \dxx + {\scriptstyle 2} \frac{u'
}{u''} \dxx - {\scriptstyle 2} \x^2 \dx2$ &\cr
\omit & height2pt & \omit && \omit &\cr
\tablerule
\omit & height2pt & \omit && \omit &\cr
&& 6 && $-\frac12 \dup \dupp - {\scriptstyle 5} \dxx - \delu \dupp \dxx - 
\frac{2}{u''} \dup + {\scriptstyle 4} \frac{u'}{u''} \dxx$ &\cr
\omit & height2pt & \omit && \omit &\cr
\tablerule
\omit & height2pt & \omit && \omit &\cr
&& 7 && $- \frac{2}{u''} - {\scriptstyle 2} \frac{x^2}{u''} \dxx$ &\cr
\omit & height2pt & \omit && \omit &\cr
\tablerule
\omit & height2pt & \omit && \omit &\cr
&& 8 && $-{\scriptstyle 5} \deldu + \delu \dup \dupp - \frac{20}{u''} - 
{\scriptstyle 4} \frac{u'}{u''} \dup - {\scriptstyle 2} \x^2 \dupp \dxx - 
{\scriptstyle 8} \frac{x^2}{u''} \dxx$ 
&\cr
\omit & height2pt & \omit && \omit &\cr
\tablerule
\omit & height2pt & \omit && \omit &\cr
&& 9 && $-\frac12 \dup \dxx - \delu \dx2$ &\cr
\omit & height2pt & \omit && \omit &\cr
\tablerule
\omit & height2pt & \omit && \omit &\cr
&& 10 && $-\frac32 \dxx - \frac12 \frac1{u''} \dup + \frac{u'}{u''} \dxx$ &\cr
\omit & height2pt & \omit && \omit &\cr
\tablerule
\omit & height2pt & \omit && \omit &\cr
&& 11 && $-\frac12 \dup \dupp - \frac{15}2 \dxx + \frac32 \delu \dup \dxx$ &\cr
&& \omit && $-\frac12 \delu \dupp \dxx - \frac12 \frac1{u''} \dup + \frac{u'}{
u''} \dxx - {\scriptstyle 2} \x^2 \dx2$ &\cr
\omit & height2pt & \omit && \omit &\cr
\tablerule
\omit & height2pt & \omit && \omit &\cr
&& 12 && $\frac52 \dup - \frac12 \dupp + \frac14 \delu \dup \dupp - \frac{5}{
u''} - \frac{u'}{u''} \dup + \frac12 \x^2 \dup \dxx - {\scriptstyle 2} 
\frac{x^2}{u''} \dxx$ &\cr
\omit & height2pt & \omit && \omit &\cr
\tablerule
\omit & height2pt & \omit && \omit &\cr
&& 13 && $0$ &\cr
\omit & height2pt & \omit && \omit &\cr
\tablerule
\omit & height2pt & \omit && \omit &\cr
&& 14 && $0$ &\cr
\omit & height2pt & \omit && \omit &\cr
\tablerule
\omit & height2pt & \omit && \omit &\cr
&& 15 && $\frac12 \dup \dupp + {\scriptstyle 2} \dxx + \delu \deldu \dxx - 
{\scriptstyle 2} \delu^2 \dx2 + \frac12 \frac1{u''} \dup - \frac{u'}{u''} 
\dxx$ &\cr
\omit & height2pt & \omit && \omit &\cr
\tablerule
\omit & height2pt & \omit && \omit &\cr
&& 16 && $0$ &\cr
\omit & height2pt & \omit && \omit &\cr
\tablerule
\omit & height2pt & \omit && \omit &\cr
&& 17 && $-{\scriptstyle 2} \dup + {\scriptstyle 5} \dupp - \delu \dup \dupp -
{\scriptstyle 14} \delu \dxx$ &\cr
&& \omit && $+{\scriptstyle 2} \delu^2 \dup \dxx + \frac{5}{u''} + \frac{u'}{
u''} \dup + {\scriptstyle 2} \x^2 \dupp \dxx - {\scriptstyle 4} \delu \x^2 \dx2
+ {\scriptstyle 2} \frac{x^2}{u''} \dxx$ &\cr
\omit & height2pt & \omit && \omit &\cr
\tablerule
\omit & height2pt & \omit && \omit &\cr
&& 18 && $-{\scriptstyle 2} \dup + \frac12 \dupp - \frac14 \delu \dup \dupp +
\frac{13}2 {\scriptstyle u'} \dxx - \frac12 \delu^2 \dupp \dxx$ &\cr
&& \omit && $+\frac12 \frac1{u''} + \frac12 \frac{u'}{u''} \dup - \frac{{u'}^2
}{u''} \dxx - \frac{11}2 {\scriptstyle u''} \dxx - \frac12 \x^2 \dup \dxx - 
\delu \x^2 \dx2$ &\cr
\omit & height2pt & \omit && \omit &\cr
\tablerule
\omit & height2pt & \omit && \omit &\cr
&& 19 && $-{\scriptstyle 4} - \frac12 \delu \dupp + \frac{u'}{u''} - 
{\scriptstyle 2} \x^2 \dxx - \frac12 \frac{x^2}{u''} \dup + \frac{u'}{u''} \x^2
\dxx$ &\cr
\omit & height2pt & \omit && \omit &\cr
\tablerule
\omit & height2pt & \omit && \omit &\cr
&& 20 && $-{\scriptstyle 24} - \frac{11}2 {\scriptstyle u'} \dup - 
{\scriptstyle 3} \delu \dupp + \frac12 \delu^2 \dup \dupp + {\scriptstyle 6}
\frac{u'}{u''} + \frac{{u'}^2}{u''} \dup + \frac92 {\scriptstyle u''} \dup$ &\cr
&& \omit && $-{\scriptstyle 16} \x^2 \dxx - \delu \deldu \x^2 \dxx + 
{\scriptstyle 2} \frac{u'}{u''} \x^2 \dxx - {\scriptstyle 2} \x^4 \dx2$ &\cr
\omit & height2pt & \omit && \omit &\cr
\tablerule
\omit & height2pt & \omit && \omit &\cr
&& 21 && $\x^2 \dup - {\scriptstyle 7} \frac{x^2}{u''} - \frac{u'}{u''} \x^2
\dup - {\scriptstyle 2} \frac{x^4}{u''} \dxx$ &\cr
\omit & height2pt & \omit && \omit &\cr
\tablerule}}

{\bf Table~5d:} {\ninepoint Operator coefficients of $\x_{\mu} t_{\nu}$ in the 
Ward identity. Act each operator on the corresponding
\vskip -8pt \noindent \hglue 3.10 truecm coefficient function in Tables~4a and 
4b, and then sum the results.}

\vfill\eject

\vskip 1cm
\vbox{\tabskip=0pt \offinterlineskip
\def\tablerule{\noalign{\hrule}}
\halign to470pt {\strut#& \vrule#\tabskip=1em plus2em& \hfil#& \vrule#& 
\hfil#\hfil& \vrule#\tabskip=0pt\cr
\tablerule
\omit & height2pt & \omit && \omit &\cr
&& \omit\hidewidth {\rm \#}\hidewidth&& \omit\hidewidth {\rm Operator}
\hidewidth&\cr
\omit & height2pt & \omit && \omit &\cr
\tablerule
\omit & height2pt & \omit && \omit &\cr
&& 1 && $-{\scriptstyle 4} \dx2$ &\cr
\omit & height2pt & \omit && \omit &\cr
\tablerule
\omit & height2pt & \omit && \omit &\cr
&& 2 && $-{\scriptstyle 2} \dx2$ &\cr
\omit & height2pt & \omit && \omit &\cr
\tablerule
\omit & height2pt & \omit && \omit &\cr
&& 3 && $0$ &\cr
\omit & height2pt & \omit && \omit &\cr
\tablerule
\omit & height2pt & \omit && \omit &\cr
&& 4 && $0$ &\cr
\omit & height2pt & \omit && \omit &\cr
\tablerule
\omit & height2pt & \omit && \omit &\cr
&& 5 && $\dupp \dxx - {\scriptstyle 2} \delu \dx2$ &\cr
\omit & height2pt & \omit && \omit &\cr
\tablerule
\omit & height2pt & \omit && \omit &\cr
&& 6 && $-\dup \dxx - {\scriptstyle 2} \delu \dx2$ &\cr
\omit & height2pt & \omit && \omit &\cr
\tablerule
\omit & height2pt & \omit && \omit &\cr
&& 7 && $-{\scriptstyle 14} \dxx - {\scriptstyle 2} \delu \dupp \dxx - 
{\scriptstyle 4} \x^2 \dx2$ &\cr
\omit & height2pt & \omit && \omit &\cr
\tablerule
\omit & height2pt & \omit && \omit &\cr
&& 8 && $-{\scriptstyle 14} \dxx + {\scriptstyle 2} \delu \dup \dxx - 
{\scriptstyle 4} \x^2 \dx2$ &\cr
\omit & height2pt & \omit && \omit &\cr
\tablerule
\omit & height2pt & \omit && \omit &\cr
&& 9 && $0$ &\cr
\omit & height2pt & \omit && \omit &\cr
\tablerule
\omit & height2pt & \omit && \omit &\cr
&& 10 && $-\frac12 \dup \dxx - \delu \dx2$ &\cr
\omit & height2pt & \omit && \omit &\cr
\tablerule
\omit & height2pt & \omit && \omit &\cr
&& 11 && $\frac12 \dupp \dxx - \delu \dx2$ &\cr
\omit & height2pt & \omit && \omit &\cr
\tablerule
\omit & height2pt & \omit && \omit &\cr
&& 12 && $-\frac14 \dup \dupp - {\scriptstyle 11} \dxx - \delu \deldu \dxx -
{\scriptstyle 3} \x^2 \dx2$ &\cr
\omit & height2pt & \omit && \omit &\cr
\tablerule
\omit & height2pt & \omit && \omit &\cr
&& 13 && $0$ &\cr
\omit & height2pt & \omit && \omit &\cr
\tablerule
\omit & height2pt & \omit && \omit &\cr
&& 14 && $0$ &\cr
\omit & height2pt & \omit && \omit &\cr
\tablerule
\omit & height2pt & \omit && \omit &\cr
&& 15 && $0$ &\cr
\omit & height2pt & \omit && \omit &\cr
\tablerule
\omit & height2pt & \omit && \omit &\cr
&& 16 && $0$ &\cr
\omit & height2pt & \omit && \omit &\cr
\tablerule
\omit & height2pt & \omit && \omit &\cr
&& 17 && $0$ &\cr
\omit & height2pt & \omit && \omit &\cr
\tablerule
\omit & height2pt & \omit && \omit &\cr
&& 18 && $\frac14 \dup \dupp + \frac12 \dxx + \frac12 \delu \deldu \dxx - 
\delu^2 \dx2$ &\cr
\omit & height2pt & \omit && \omit &\cr
\tablerule
\omit & height2pt & \omit && \omit &\cr
&& 19 && $-{\scriptstyle 3} \dup + \frac12 \dupp - \frac12 \delu \dup \dupp -
{\scriptstyle 8} \delu \dxx - \delu^2 \dupp \dxx$ &\cr
&& \omit && $-\x^2 \dup \dxx - {\scriptstyle 2} \delu \x^2 \dx2$ &\cr
\omit & height2pt & \omit && \omit &\cr
\tablerule
\omit & height2pt & \omit && \omit &\cr
&& 20 && $-\frac12 \dup + {\scriptstyle 3} \dupp - \frac12 \delu \dup \dupp
-{\scriptstyle 8} \delu \dxx + \delu^2 \dup \dxx$ &\cr
&& \omit && $+\dupp \x^2 \dxx - {\scriptstyle 2} \delu \x^2 \dx2$ &\cr
\omit & height2pt & \omit && \omit &\cr
\tablerule
\omit & height2pt & \omit && \omit &\cr
&& 21 && $-{\scriptstyle 42} - {\scriptstyle 7} \delu \deldu + \delu^2 \dup 
\dupp - {\scriptstyle 30} \x^2 \dxx$ &\cr
&& \omit && $-{\scriptstyle 2} \delu \deldu \x^2 \dxx - {\scriptstyle 4} \x^4 
\dx2$ &\cr
\omit & height2pt & \omit && \omit &\cr
\tablerule}}

{\bf Table~5e:} {\ninepoint Operator coefficients of $\x_{\mu} x_{\nu}$ in the 
Ward identity. Act each operator on the corresponding
\vskip -8pt \noindent \hglue 3.10 truecm coefficient function in Tables~4a and 
4b, and then sum the results.}

\vskip .5cm
\centerline{V. DISCUSSION}

We have computed one loop graviton self-energy in a locally de Sitter 
background and using the integral approximations (2.19a-b) for the 
pseudo-graviton and ghost propagators. The result consists of 21 independent 
tensors, given in Table~3, times the coefficient functions of Tables~4a and 4b.
Not the least of our conclusions is that the result is almost certainly 
correct, within the integral approximation. It is difficult to doubt this as 
one witnesses the cancellation of one after another of the hundreds of distinct
functional and tensor terms in the Ward identity. Agreement with Capper's flat
space result$^7$ shows that we have even got the sign and the normalization 
right.

A subtle and interesting point is that the forms (2.19a-b) used for the 
pseudo-graviton and ghost propagators are only approximations to the exact 
mode expansions which one obtains on $T^3 \times \Re$.$^8$ The integral 
approximations
become exact in the flat space limit, so it is obvious why the flat space limit
of our result should agree with Capper's work. The fascinating thing is that 
the Ward identity is also obeyed, exactly and without any need for taking the 
flat space limit. We saw this as well when checking a somewhat more general Ward
identity at tree order.$^{11}$ The reason for it seems to be that the integral
approximations do invert the pseudo-graviton and ghost kinetic operators, so
they differ from the true propagators only by real, analytic terms which depend
upon the choice of vacuum. There can only be mixing with these vacuum dependent
terms if the Ward identity involves integrations which can reach the initial or
final states. But the one loop identity we checked involves no integrations at 
all.

Although the one loop self-energy is worthy of study in its own right, our 
interest derives from its role as an important constituent in the two loop
tadpole from which we have lately inferred the quantum gravitational back
reaction on inflation.$^4$ Of course the one loop self-energy cannot completely
verify the two loop tadpole, but it does establish the correctness of certain 
features of the basic formalism. For example, our gauge fixing procedure is 
shown to be consistent, and the  $\psi^3$ and the $\psi {\overline \omega} 
\omega$ vertex operators are checked. The result also demonstrates the validity
of our tensor contraction routines, and the procedures whereby derivative 
operators from the vertices are acted on propagators. Since the same vertices 
and reduction procedures were used throughout the two loop work, many features 
of the larger calculation are checked as well. For example, the diagrams in 
Fig.~1a and Fig.~1b are obtained by contracting the one loop self-energy, 
through two pseudo-graviton propagators, into the $\psi^3$ vertex operator, as
illustrated in Fig.~2. No new vertices or propagators appear, and the same 
procedures were used to perform the contractions and to act the derivatives. 
These comments apply as well to the outer ghost loop of Fig.~1c, and to 
everything but the 4-point vertex operator in the 4--3 diagram of Fig.~1d. One 
point the current work does not check is the procedures for integrating over 
free interaction vertices. Of course the integrals have been checked 
extensively in other ways,$^4$ but not by the one loop self-energy.

\vskip .5cm
\centerline{ACKNOWLEDGEMENTS}

We wish to thank T. J. M. Zouros for the use of his SUN SPARC station, and S.
Deser for his encouragement and support during this difficult project. One of
us (RPW) thanks the University of Crete and the Theory Group of FO.R.T.H. for
their hospitality during the execution of this project. This work was partially
Supported by DOE contract DE-FG05-86-ER40272, by NATO Collaborative Research
Grant CRG-910627, and by NSF grant INT-94092715.

\vskip .5cm
\centerline{REFERENCES}
\item{1.} N. C. Tsamis and R. P. Woodard, {\sl Phys. Lett.} {\bf B301} (1993)
351.

\item{2.} N. C. Tsamis and R. P. Woodard, {\sl Ann. Phys.} {\bf 238} (1995) 1.

\item{3.} N. C. Tsamis and R. P. Woodard, {\sl Nucl. Phys.} {\bf B474} (1996)
235.

\item{4.} N. C. Tsamis and R. P. Woodard, The Quantum Gravitational 
Back-Reaction On Inflation,'' {\it hep-ph/9602316}. (To appear in {\sl Annals
of Physics}.)

\item{5.} M. Goroff and A. Sagnotti, {\sl Phys. Lett.} {\bf B160} (1986) 81; 
{\sl Nucl. Phys.} {\bf B266} (1986) 709.

\item{6.} J. Schwinger, {\sl J. Math. Phys.} {\bf 2} (1961) 407; {\it 
Particles, Sources and Fields} (Addison-Wesley, Reading, MA, 1970).

\item{7.} D. M. Capper, {\sl J. Phys.} {\bf A13} (1980) 199.

\item{8.} N. C. Tsamis and R. P. Woodard, {\sl Class. Quantum Grav.} {\bf 11}
(1994) 2969.

\item{9.} N. C. Tsamis and R. P. Woodard, {\sl Commun. Math. Phys.} {\bf 162}
(1994) 217.

\item{10.} S. Deser and L. F. Abbott, {\sl Nucl. Phys.} {\bf B195} (1982) 76.
\hfill\break P. Ginsparg and M. J. Perry, {\sl Nucl. Phys.} {\bf B222} (1983) 
245.

\item{11.} N. C. Tsamis and R. P. Woodard, {\sl Phys. Lett.} {\bf B292} (1992)
269. 

\item{12.} B. S. DeWitt, {\sl Phys. Rev.} {\bf 162} (1967) 1239.\hfill\break
F. A. Berends and R. Gastmans, {\sl Nucl. Phys.} {\bf B88} (1975) 99.

\item{13.} S. Wolfram, {\it Mathematica, 2nd Edition} (Addison-Wesley, Redwood
City, CA, 1991).

\item{14.} R. Mertig, {\it Guide to FeynCalc 1.0}, University of W\"urzburg 
preprint, March 1992.

\item{15.} D. M. Capper, private communication of July 29, 1982.

\bye